\documentclass[aps,jcp,preprint,onecolumn,floatfix,nofootinbib,superscriptaddress]{revtex4-1}

\usepackage{amsmath}
\usepackage{amssymb}
\usepackage{amsfonts}
\usepackage{float}
\usepackage{mathtools}
\usepackage{listings}
\usepackage{bm}

\newcommand{\quotes}[1]{``#1''}
\newcommand{\bra}[1]{\ensuremath{\langle \: #1 \: |}}
\newcommand{\ket}[1]{\ensuremath{| \: #1 \: \rangle}}
\newcommand{\OvI}[2]{\ensuremath{\langle \: #1 \mid #2 \: \rangle}}
\newcommand{\ExV}[3]{\ensuremath{\langle \: #1 \mid #2 \mid #3 \: \rangle}}

\begin{document}

\pagenumbering{arabic}

\title{Nuclear-Electronic All-Particle\\ Density Matrix Renormalization Group}

\author{Andrea Muolo}
\affiliation{ETH Z\"urich, Laboratory of Physical Chemistry, Vladimir-Prelog-Weg 2, 8093 Z\"urich, Switzerland}

\author{Alberto Baiardi}
\affiliation{ETH Z\"urich, Laboratory of Physical Chemistry, Vladimir-Prelog-Weg 2, 8093 Z\"urich, Switzerland}

\author{Robin Feldmann}
\affiliation{ETH Z\"urich, Laboratory of Physical Chemistry, Vladimir-Prelog-Weg 2, 8093 Z\"urich, Switzerland}

\author{Markus Reiher}
\email{corresponding author: markus.reiher@phys.chem.ethz.ch}
\affiliation{ETH Z\"urich, Laboratory of Physical Chemistry, Vladimir-Prelog-Weg 2, 8093 Z\"urich, Switzerland}

\begin{abstract}

We introduce the Nuclear Electronic All-Particle Density Matrix Renormalization Group (NEAP-DMRG) method 
for solving the time-independent Schr\"odinger equation simultaneously for electrons
and other quantum species. 
In contrast to already existing multicomponent approaches, 
in this work we construct from the outset a multi-reference trial wave function
with stochastically optimized non-orthogonal Gaussian orbitals.
By iterative refining of the Gaussians' positions and widths, 
we obtain a compact multi-reference expansion for the multicomponent wave function.
We extend the DMRG algorithm to multicomponent wave functions to take into account inter- and intra-species correlation effects.
The efficient parametrization of the total wave function as a matrix product state allows NEAP-DMRG to
accurately approximate full configuration interaction energies of molecular systems with more than 
three nuclei and twelve particles in total, which is currently a major challenge for other multicomponent approaches.
We present NEAP-DMRG results for two few-body systems, i.e., H$_2$ and H$_3^+$, and one larger system, namely BH$_3$.

\end{abstract}

\maketitle

\section{Introduction}
\label{SEC:intro}

Modeling the static and dynamical properties of molecular systems relies routinely on the Born--Oppenheimer (BO) approximation \cite{Born-Oppenheimer1927,BornHuang1954} which separates 
the time-independent Schr\"odinger equation into the electronic and nuclear parts,
motivated by the large mass difference between the two subsystem components. 
Systematic corrections to this approximation can be considered \textit{a posteriori} 
by non-adiabatic terms that couple the various potential energy surfaces (PESs), 
which are also an outcome of the BO approximation.
Although such an \textit{a-posteriori} approximation of non-adiabatic couplings 
has been well developed and delivers highly accurate results, 
it can produce large errors when nuclear quantum effects are not negligible \cite{Validity_BOA}.
The BO approximation fails when PESs of different electronic states are close to each other or even exhibit conical intersections \cite{Cederbaum1984}.
This is especially evident in excited-state dynamics \cite{QuantumCoherence1,QuantumCoherence2,QuantumCoherence3,QuantumCoherence4}, 
e.g. in photochemical systems \cite{photosynthesis1,photosynthesis2,vision1,vision2,vision3}
and for charge- and proton-transfer reactions  \cite{hydrogenstorage1,hydrogenstorage2,hydrogenstorage3,hydrogenstorage4,photovoltaic1,photovoltaic2,photovoltaic3}.

The calculation of a nuclear wave function relies on the availability of 
a PES itself, which is determined by interpolating point-wise defined electronic energies, obtained as
solutions of the electronic Schr\"{o}dinger equation for different nuclear configurations.
Despite the recent development of efficient schemes to obtain high-dimensional PESs based 
on permutationally invariant polynomials \cite{Bowman2018_PIP-Review}, 
machine learning algorithms \cite{Behler2016_MLPES-Review,Paesani2016_WaterPotential-Review,Paesani2018_WaterPotential,Bowman2018_GaussianProcesses}, 
and neural networks \cite{Parrinello2007_PES-NN,Behler2007_PES-NN,Manzhos2018_NNvsML,Brorsen2019_NeuralNetwork}, 
approximating global PESs that accurately describe the energy in the dissociative limit
still represents a major challenge for systems with more than five to six nuclei \cite{Dawes2012_CO2-Global,Dawes2012_GlobalPES-Ozone,Truhlar2013_PES-N4}.
Furthermore, the BO approximation introduces large errors in calculations for more exotic systems 
comprising particles that are electron-like but heavier or nuclear-like but lighters such as muons ($\mu^{\pm}$) and positrons ($e^+$), 
for which theoretical approaches are continuously being developed \cite{Blundell2015_MuonicDFT,lind18,Shahbazian2018_Muonic-MP2}.

So-called multicomponent methods attempt to bypass the limitations outlined above 
by solving the full time-independent Schr\"{o}dinger equation for electrons and other particles without invoking the BO approximation.
Explicitly correlated basis functions, which cannot be factorized into single-particle functions
as they explicitly depend on every inter-particle distance,
paved the way toward highly accurate solutions of the full time-independent Schr\"{o}dinger equation.
Despite their successes, these methods are limited to few-body systems, up to LiH
\cite{Adamowicz2015_LiH}, due to the factorial scaling
with the number of particles when imposing the correct permutational symmetry \cite{Varga1995,Korobov2000,Matyus2012,Adamowicz2012,Adamowicz2013_rev,Cox-4,Pachucki2016,muolo2019}.
Note, however, that such explicitly correlated functions are the key to accurate electronic-structure calculations where they,
as Jastrow factors, alleviate the cusp-condition peculiarities of Gaussian orbitals \cite{Klopper2006_R12-Review,Ten-no2012_R12-Wires,Valeev2012_R12-Review}.

Parallel to the development of explicitly correlated methods 
and in analogy to one-electron orbital-based approaches,
the introduction of nuclear orbitals led to the extension
of well-established (post-)Hartree--Fock (HF) methods to the mixed nuclear-electron problem,
which allow for the study of
larger systems than those accessible with explicitly correlated variants.
Multicomponent orbital-based methods proposed in the last two decades
construct orbitals as solutions of the mean-field self-consistent-field (SCF) nuclear-electronic HF equations;
these are the nuclear orbital plus molecular orbital (NOMO) 
\cite{Nakai1998_NOMO-Original,nakai2002simultaneous,Nakai2003,nakai2007nuclear},
the nuclear electronic orbital (NEO) \cite{Hammes-Schiffer2002,swalina2005alternative,Hammes-Schiffer2005_NOCI, Hammes-Schiffer2008_Geminals,Hammes-Schiffer2011_ExplicityCorrelated},
the multicomponent molecular orbital (MCMO)\cite{Tachikawa1998,Tachikawa2002}, and
the any-particle molecular orbital (APMPO)\cite{APMO_2008} approaches.

In addition to the electron-electron correlation, 
any multicomponent mean-field approach must include 
both the nucleus-nucleus and the nucleus-electron correlations to obtain an accurate representation of the wave function.
The latter class of interactions is particularly challenging to describe in terms of one-particle functions, i.e., orbitals, 
because the electrons adapt quickly to the comparatively slow dynamics of the nuclei.
As a result, HF energies and single-reference wave functions 
can be qualitatively inaccurate~\cite{HammesSchiffer2004_Tunneling-Correlation,Nakai2011_ECGplusNOMO}.
To lift this limitation, several post-HF schemes have been developed, 
including multicomponent coupled cluster (CC)~\cite{Chakraborty2016_multicomponent_CC,Pavosevic2019_MulticomponentCC}, 
configuration interaction (CI)~\cite{Hammes-Schiffer2005_NOCI,Patrick_2015_EN-MFCI,Patrick_2017_EN-MCFI},
density functional theory (DFT)~\cite{Hammes-Schiffer2012_MulticomponentDFT,Hammes-Schiffer2017_MulticomponentDFT,Hammes-Schiffer2017_PracticalMulticomponentDFT,CoupledMeanFieldTheory},
and Green's function methods \cite{Reyes2014_particlepropagatortheory}.
Multicomponent DFT exchange-correlation functionals inspired by the Colle-Salvetti formalism can reproduce qualitatively correct proton energies and densities of small molecules \cite{Brorsen2018_TransferableDFT,Yang2018_Stability,HammesSchiffer2019_GGA-DFT}.
Nevertheless, a transferable multicomponent functional that is accurate for a wide range of systems has not been available yet.
Moreover, the design of new DFT functionals requires highly-accurate benchmark data obtained from post-HF methods.
However, the scaling of methods such as CC and CI with respect to system size impedes their applications to large molecules.

Even if the vast majority of multicomponent schemes proposed so far
considers a single-reference wave function,
a nuclear-electronic wave function can be multi-reference due to 
the strong correlation between the electrons and the nuclei~\cite{HammesSchiffer2004_Tunneling-Correlation}.
This effect has been highlighted by Brorsen~\cite{Brorsen2020_SelectedCI-PreBO} 
who extended the heath-bath CI algorithm~\cite{Urmigar2016_HBCI}
to multicomponent systems.
In the present work, we therefore account for the multicomponent correlation problem in two ways
different from the aforementioned approaches.
(i) We introduce a stochastic multireference-multicomponent trial wave function, 
which is composed of properly (anti-)symmetrized products of 
non-orthogonal nuclear-electronic all-particle (NEAP) functions. 
The NEAP function's parameters, e.g. the Gaussian shifts and widths, 
are (re)optimized variationally in order to minimize the length of the multireference expansion.
(ii) We then present a multicomponent extension to the density matrix renormalization  group (DMRG) algorithm
\cite{White1992,white1993density,Hastings2007_AreaLaw}.
to optimize the resulting multireference--multicomponent wave function
Based on the so-called second-generation formulation of DMRG, we encode the multicomponent CI wave function as a matrix product state (MPS) \cite{Oestlund1995_MPS} and introduce a new algorithm to express any multicomponent Hamiltonian as a matrix product operator (MPO) \cite{keller2015efficient} starting from its second-quantization form \cite{streltsov2010general}.
Note that another multicomponent DMRG variant has recently been introduced by Yang and White
\cite{White2019_PreBO} to study the H$_2$ and Ps$_2$ molecules. 
However, they rely on a one-dimensional approximation of the three-dimensional
real-space representation of the full molecular Hamiltonian that
is limited to diatomic molecules. By contrast, our NEAP-DMRG method can be applied
to any molecule since it relies on the second-quantization representation 
of the full molecular Hamiltonian obtained with an arbitrary orbital basis.

Our work is organized as follows. Section~II introduces the theoretical framework.
First, the translation-free Hamiltonian and the wave function ansatz are introduced.
Next, we propose the multistage variational optimization of the NEAP orbitals.
Section~III starts with defining the multicomponent full-CI space which is followed
by the extension of the DMRG algorithm to multicomponent wave functions and Hamiltonians.
Section~IV presents numerical results of our NEAP-DMRG method 
for H$_2$, H$_3^+$, and BH$_3$. Concluding remarks are given in Section~V.

\section{Theory}
\label{sec:theory}

\subsection{Translation-free pre-Born--Oppenheimer Hamiltonian}

Given $N_{\text{p}}$ particles with masses $m_\mu$ and electric charges $q_\mu$, we define 
the collective position vector expressed in laboratory-fixed Cartesian coordinates as $\mathbf{r}=(\mathbf{r}_1,\ldots,\mathbf{r}_{N_{\text{p}}})^T$. 
The non-relativistic Schr\"odinger Hamiltonian in Hartree atomic units reads
\begin{equation} 
 \mathcal{H} = -\sum_{\mu}^{N_{\text{p}}} \frac{1}{2 m_\mu} \bm{\nabla}^2_\mu
 +\sum_{\mu<\nu}^{N_{\text{p}}} \frac{q_\mu q_\nu}{|\mathbf{r}_\mu -\mathbf{r}_\nu|} ~,
 \label{eq:PreBO-Hamiltonian}
\end{equation}
where $\bm{\nabla}_{\mu}=(\nabla_{\mu x},\nabla_{\mu y},\nabla_{\mu z})^T$ is the derivative operator with respect to the $\mu$-th particle's Cartesian coordinates.
We first remove from the Hamiltonian of Eq.~(\ref{eq:PreBO-Hamiltonian}) the unwanted 
center-of-mass (CM) translational contributions
that would lead to a set of continuum-energy states \cite{Nakai2005_TRF}. 
To this end, we consider a linear transformation of the coordinates, 
\begin{equation}
 U_x \mathbf{r} = (\mathbf{x}_1, \ldots, \mathbf{x}_{N_{\text{p}-1}}, \mathbf{x}_{\text{CM}})^T \, ,
 \label{eq:CoordinateTransformation}
\end{equation}
in which the CM position, $\mathbf{x}_{\text{CM}}=\sum_{\mu}^{N_{\text{p}}} m_\mu \mathbf{r}_\mu/\big(\sum_{\mu}^{N_{\text{p}}} m_\mu\big)$, is separated from the remaining $N_{\text{p}}-1$ translationally invariant positions. 
The (infinitely many) linear transformations $U_x$ that separate the CM coordinates 
as in Eq.~(\ref{eq:CoordinateTransformation}), must satisfy the following conditions
\begin{eqnarray}
 \label{eq:TranslationalInvariance1}
  \sum_{\nu}^{N_{\text{p}}} (U_x)_{\mu\nu} &=&0 
    \hspace{0.5cm} {\text{with}} \hspace{0.5cm} \mu=1, ... N_{\text{p}}-1 ~, \\
                 (U_x)_{N_{\text{p}}\nu} &=&\frac{m_\nu}{M} \, ,
 \label{eq:TranslationalInvariance2}
\end{eqnarray}
with $M=\sum_{\mu}^{N_{\text{p}}} m_\mu$. 
The CM translational energy contamination of the total energy is given by the CM kinetic energy term
\begin{equation}
  \mathcal{T}_{\text{CM}} = -\bm{\nabla}^T_{\text{CM}} \frac{1}{2M} \bm{\nabla}_{\text{CM}} ~,
  \label{eq:CMKineticEnergys}
\end{equation}
that can be transformed to laboratory-fixed Cartesian coordinates 
\begin{equation}
  \mathcal{T}_{\text{CM}} =
    -\sum_{\mu\nu}^{N_{\text{p}}}  \bm{\nabla}^T_{\text{CM}} 
  (U^T)_{\nu N_{\text{p}}} (U^{-T})_{N_{\text{p}}\nu} \frac{1}{2M} 
    (U^{-1})_{\mu N_{\text{p}}} (U)_{N_{\text{p}}\mu} \bm{\nabla}_{\text{CM}} 
    = -\sum_{\mu\nu}^{N_{\text{p}}}  \bm{\nabla}_\nu \frac{1}{2M} \bm{\nabla}_\mu
 \label{eq:CMKineticEnergy2}
\end{equation}
where we exploit the relation $(U^{-1})_{\mu N_{\text{p}}}=1$, proved in our previous work \cite{muolo2018generalized}, 
and where we take into account that $(U_x)_{\mu\nu}=\partial x_\mu/\partial r_\nu$. Finally, the translation-free Hamiltonian is obtained by subtracting Eq.~(\ref{eq:CMKineticEnergy2})
from Eq.~(\ref{eq:PreBO-Hamiltonian})
\begin{equation}
  \mathcal{H}_{\text{TF}} =
   - \sum_{\mu}^{N_{\text{p}}} \left(\frac{1}{2 m_\mu}-\frac{1}{2M}\right) \bm{\nabla}^2_\mu
   + \sum_{\mu<\nu}^{N_{\text{p}}} \frac{q_\mu q_\nu}{|\mathbf{r}_\mu -\mathbf{r}_\nu|} 
   + \sum_{\mu<\nu}^{N_{\text{p}}} \frac{1}{M} \bm{\nabla}_\mu\bm{\nabla}_\nu ~.
  \label{eq:TranslationallyInvariantHam}
\end{equation}
Some authors work out the same elimination of the CM degree of freedom but
neglect the electronic contribution to the translational motion,
which is believed to be negligible compared to the nuclear contributions
\cite{Nakai2011_ECGplusNOMO}.
Although this assumption is justified by the mass difference between the two subsystem components,
we consider here the exact translation-free Hamiltonian in Eq.~(\ref{eq:TranslationallyInvariantHam}).
We highlight that the latter Hamiltonian contains only one- and two-body terms, and therefore, it is of
the same complexity as the non-relativistic electronic Hamiltonian, the major difference being
the presence of quantum particles of different types. 
The presence of only one- and two-body terms makes the translation-free Hamiltonian of Eq.~(\ref{eq:TranslationallyInvariantHam}) potentially
simpler than the BO vibrational Hamiltonian that contains, in general, arbitrary-order coupling terms
in the potential \cite{Kongsted2006_NMode}.

\subsection{Multicomponent Wave-Function Ansatz}
\label{subsec:wf}

In analogy with multireference methods in electronic-structure calculations, where the approximate electronic wave function is expanded in terms of Slater determinants, we represent the exact wave function $\Psi$ of the complete molecular system (including both nuclei and electrons), as a (finite) linear combination of $N_b$ independent non-orthogonal basis functions $\psi_I$
\begin{equation}
 \Psi = \sum_I^{N_{\text{b}}} c_I \psi_I ~. 
 \label{eq:GeneralizedCI}
\end{equation}
Each total-system basis function $\psi_I$ is a product of (electronic or nuclear) properly (anti-)symmetrized many-body functions $\Phi_{Ii}$
\begin{equation}
 \psi_I = \prod_i^{N_{\text{t}}} \Phi_{Ii} = 
   \prod_i^{N_{\text{t}}} \left( \mathcal{S}^{(i)}_{\pm1} 
   \prod_\mu^{N_{\text{p}}^{(i)}} {\phi}_{Ii\mu}^{s_i m_{s_{i\mu}}} \right) ~,
 \label{eq:conf}
\end{equation}
where $N_t$ is the number of the different particle types of the molecule,
e.g. $N_t=2$ for H$_2$ and $N_t=3$ for H$_2$O. 
$N_p^{(i)}$ is the number of particles of a given type $i$ 
(note that this formalism is easily extendable to various types of particles other than electrons and atomic nuclei).
$\Phi_{Ii}$ is a product of orbitals (anti-)symmetrized by $\mathcal{S}^{(i)}_{\pm1}$, 
the (anti-)symmetrization operator:
\begin{equation}
 \mathcal{S}^{(i)}_{\pm} = 
 \begin{cases}
   \mathcal{S}^{(i)}_{+}  = \frac{1}{N!} 
     \sum_{\mathcal{P} \in S_N} \mathcal{P} & \text{bosons,} \\
   \mathcal{S}^{(i)}_{-} = \frac{1}{N!} \sum_{\mathcal{P} \in S_N}
     \mathrm{sgn}(\mathcal{P}) \mathcal{P} & \text{fermions} ~,
 \end{cases}
 \label{eq:Antisymmetrization}
\end{equation}
with $S_N$ being the permutation group for $N$ identical objects. Finally, each orbital $\phi_{Ii\mu}^{s_i m_{s_{i\mu}}}$ is the product of a spatial part, $\varphi_{Ii\mu}$, and the spin function $\chi_{Ii\mu}^{s_i m_{s_{i\mu}}}$ with overall spin $s_i$ and spin projection $m_{s_{i\mu}}$:
\begin{equation}
  \phi_{Ii\mu}^{s_i m_{s_{i\mu}}} = \varphi_{Ii\mu} \, \chi_{Ii\mu}^{s_i m_{s_{i\mu}}} ~.
  \label{eq:SpatialOrbital}
\end{equation}
For the sake of simplicity, we refer to the one-particle functions as molecular orbitals in close analogy to the 
molecular orbitals in electronic structure theory
and compose $\varphi_{Ii\mu}$ for any particle as a linear combination of (nuclear or electronic) primitive orbitals (LCPO)
\begin{equation}
 \varphi_{Ii\mu}  = \sum_\nu^{N_\text{LCPO}^{(i)}} d_{Ii\mu\nu}
 ~g_{Ii\mu\nu}(\mathbf{r}_{i\mu};\alpha_{Ii\mu\nu},\mathbf{s}_{Ii\mu\nu}) \, ,
 \label{eq:LCPO}
\end{equation}
where $d_{Ii\mu\nu}$ are expansion coefficients and $g_{Ii\mu\nu}$ are Gaussian-type orbitals 
with Gaussian widths $\alpha_{Ii\mu\nu}$ and displacements $\mathbf{s}_{Ii\mu\nu}$
\begin{equation}
 g_{Ii\mu\nu}(\mathbf{r}_{i\mu};\alpha_{Ii\mu\nu},\mathbf{s}_{Ii\mu\nu}) 
 = \mathrm{exp} \left[ -\alpha_{Ii\mu\nu} 
 \left(\mathbf{r}_{i\mu} -\mathbf{s}_{Ii\mu\nu}\right)^2 \right] \, ,
 \label{eq:GaussianBasis}
\end{equation}
with $\alpha_{Ii\mu\nu} \in \mathbb{R}^+$ and $\mathbf{s}_{Ii\mu\nu} \in \mathbb{R}^3$.

From here on, we refer to the parametrization of Eq.~(\ref{eq:GeneralizedCI}) as a \textit{generalized CI expansion}. Unlike the well-known purely electronic formulation of CI, the many-body functions ${\Phi}_{Ii}$ consist of products of Slater determinants (permanents) for every set of fermionic (bosonic) species. Furthermore, the orbital basis $\{{\phi}_{Ii\mu}^{s_i m_{s_{i\mu}}}\}$ is, in general, different for each many-body function $\psi_I$. This is not true for standard CI, where a unique self-consistently optimized single-particle basis is employed to construct all Slater determinants. 
In this respect, our generalized CI expansion is very similar to a multi-reference CI parametrization \cite{Lischka2018_ExcitedStatesMR-Review}.

In agreement with the notation introduced in this paragraph, capital roman indices correspond to total-system basis functions, lower case roman indices to particle types, and lower case Greek ones to single-particle functions.

\subsection{Expectation Values}
\label{SUBSEC:matel}

The expectation value of an operator $\mathcal{O}$ over a wave function $\Psi$ of Eq.~(\ref{eq:GeneralizedCI}) reads
\begin{equation}
  \langle \mathcal{O} \rangle  
    = \frac{ \ExV{\Psi}{\mathcal{O}}{\Psi} }{ \OvI{\Psi}{\Psi}}
    = \frac{\sum_{IJ}^{N_{\text{b}}} c^*_I c_J \prod_{ij}^{N_t}\ExV{\Phi_{Ii}}{\mathcal{O}}{\Phi_{Jj}} } 
           {\sum_{IJ}^{N_{\text{b}}} c^*_I c_J \prod_{i}^{N_{\text{t}}} \OvI{\Phi_{Ii}}{\Phi_{Ji}} } ~.
  \label{eq:ExpectationValue}
\end{equation}

The operator $\mathcal{O}$ can be represented in the basis set $\{\psi_I\}$,
${\bm{O}}=\{O_{IJ} \}$, with the ($I,J$)-th matrix element given by
\begin{equation}
 O_{IJ} = \ExV{ \psi_I }{ \mathcal{O} }{ \psi_{J} }
        = \prod_{ij}^{N_{\text{t}}} \ExV{ \Phi_{Ii} }{ \mathcal{O} }{ \Phi_{Jj} } ~.
 \label{eq:MatrixRepresentation}
\end{equation}

Eq.~(\ref{eq:ExpectationValue}) can be written compactly in matrix notation by collecting the coefficients entering the linear combination in a vector $\mathbf{c} = \left( c_1,\ldots,c_{N_\text{b}} \right)^T$
\begin{equation}
 \langle \mathcal{O} \rangle = \frac{\mathbf{c}^\mathrm{T} {\bm{O}} \mathbf{c} }
 {\mathbf{c}^\mathrm{T} {\bm{S}} \mathbf{c} } ~,
 \label{eq:CompactExpectationValue}
\end{equation}
where $\bm{S}$ is the matrix representation of the identity operator in this basis, i.e. the overlap matrix.
The matrix elements in Eq.~(\ref{eq:MatrixRepresentation}) involve integrals of quantum mechanical operators over many-body functions $\Phi_{Ii}$ that are (anti-)symmetrized products of single-particle functions $\phi_{Ii\mu}$. In Born--Oppenheimer electronic-structure calculations, the molecular orbitals are required to be orthogonal so that the evaluation of expectation values is enormously simplified by the Slater-Condon rules. 
In this work, we discard the orthogonality constraint and optimize non-orthogonal sets of electron-nuclear single-particle functions in order to increase the flexibility of the basis set and obtain a compact expansion of Eq. (\ref{eq:GeneralizedCI}), i.e., one that requires only a small $N_\text{b}$. 
Rules for calculating integrals in a non-orthogonal single-particle basis have been previously derived by L{\"o}wdin \cite{lowdin1955quantum}. We generalize his work to multicomponent systems composed of both fermionic and bosonic species and present the explicit derivation in the
next section \ref{sub:MatEl}.

\subsection{Evaluation of Matrix Elements}
\label{sub:MatEl}

The following derivation of the matrix elements between Slater determinants (permanents)
constructed of non-orthogonal one-particle orbitals, follows
L{\"o}wdin's steps \cite{lowdin1955quantum} generalized to multicomponent systems.
For the sake of brevity, we drop hereafter the subscript $I,J$ labeling the basis functions.

The (anti)symmetrized function $\Phi_{n}$ can be expanded through a Laplace expansion as follows
\begin{equation}
 \Phi_{n}
 = \mathcal{S}^{(n)}_{\pm}  \prod_\nu^{N_p^{(n)}} \phi_{n\nu}(\mathbf{r}_{n\nu})
 = \sum_\alpha^{N_p^{(n)}} \phi_{n\alpha}(\mathbf{r}_{n\lambda}) 
 (\pm 1)^{\alpha+\lambda} \Phi_{n}^\pm(\lambda|\alpha) ~,
\end{equation}
with $\Phi_{n}^\pm(\lambda|\alpha)$ being the Slater determinant (permanent) 
having subtracted the $\lambda$-th row and $\alpha$-th column.
The $IJ$ matrix element $O^{[1]}_{n\lambda}$ for a one-body operator $o_n(\lambda)$ 
acting on the $\lambda$-th particle of type $n$ reads
\begin{align}
 O^{[1]}_{n\lambda}  &=  
 \left\langle \psi \Big| o_n(\lambda) \Big|\psi\right\rangle
 = N_n \left\langle \Phi_n \Big| o_n(\lambda) \Big| \Phi_n \right\rangle
 \nonumber \\
 &= N_n  \left\langle 
 \mathcal{S}^{(n)}_{\pm} \prod_\mu^{N_p^{(n)}}
 \phi_{n\mu}(\mathbf{r}_{n\mu}) 
 \Big| o_n(\lambda) \Big| 
 \mathcal{S}^{(n)}_{\pm}  \prod_\nu^{N_p^{(n)}}    
 \phi_{n\nu}(\mathbf{r}_{n\nu})
 \right\rangle
 \nonumber \\
 &= N_n  \left\langle 
 \prod_\mu^{N_p^{(n)}}     
 \phi_{n\mu}(\mathbf{r}_{n\mu}) 
 \Big| o_n(\lambda) \Big| 
 \mathcal{S}^{(n)}_{\pm}  \prod_\nu^{N_p^{(n)}}    
 \phi_{n\nu}(\mathbf{r}_{n\nu})
 \right\rangle
 \nonumber \\
 &= N_n \sum_{\alpha}^{N_{\text{p}}^{(n)}} (\pm 1)^{\alpha+\lambda}
 \left\langle   
 \prod_{\mu}^{N_p^{(n)}} \phi_{n\mu}(\mathbf{r}_{n\mu})
 \Big| o_n(\lambda) \Big|
 \phi_{n\alpha}(\mathbf{r}_{n\lambda})          
 \Phi_{n}^\pm(\lambda|\alpha)
 \right\rangle
 \nonumber\\
 &=  N_n \sum_{\alpha}^{N_p^{(n)}}
 C^{(n)}_{\lambda,\alpha}
 \left\langle 
 \phi_{n\lambda}(\mathbf{r}_{n\lambda})
 \Big| o_n(\lambda) \Big|
 \phi_{n\alpha}(\mathbf{r}_{n\lambda})
 \right\rangle_{\mathbf{r}_{n\lambda}} ~,
 \label{eq:1bodyMatElm}
\end{align}
where in the second step, we exploited hermiticity, 
the commutativity relation $[S_{\pm},o(\lambda)]=0$, 
and the idempotency of the (anti)symmetrizer,
while $N_n$ is defined as
\begin{equation}
 N_n = \prod_{i\neq n}^{N_t} 
 \left\langle \Phi_{i} \Big{|}\Phi_{i}\right\rangle ~,
\end{equation}
and $C^{(n)}_{\lambda,\alpha}$ is the product of 
the residual overlap matrix elements
\begin{equation}
 C^{(n)}_{\lambda,\alpha} = (\pm 1)^{\alpha+\lambda}
 \left\langle \prod_{\mu\neq \lambda}^{N_p^{(n)}}
 \phi_{n\mu}(\mathbf{r}_{n\mu})
 \Bigg|
 \Phi_{n}^\pm(\lambda|\alpha)
 \right\rangle_{\forall \mathbf{r}_{n\mu\neq n\lambda}} ~.
 \label{eq:cofactor}
\end{equation}

The case of two-body operators $\mathcal{O}^{[2]}$ is slightly more involved
and requires the Laplace expansion to be applied twice
\begin{align}
 \Phi_{n} 
 &= 
 \sum_{\alpha<\beta}^{N_{\text{p}}^{(n)}}
 (1 \pm \mathcal{P}_{\lambda\kappa})
 \phi_{n\alpha}(\mathbf{r}_{n\lambda}) 
 \phi_{n\beta}(\mathbf{r}_{n\kappa})~ 
 (\pm 1)^{\alpha+\beta+\lambda+\kappa} ~
 \Phi_{n}^\pm(\lambda\kappa|\alpha\beta) ~,
\end{align}
where $\Phi_{n}^\pm(\lambda\kappa|\alpha\beta)$ is 
the cofactor function associated with the Slater determinant (permanent) having eliminated 
the $\lambda$-th and $\kappa$-th row and $\alpha$-th and $\beta$-th column. 
Lastly, $\mathcal{P}_{\lambda\kappa}$ permutes particles $\lambda$ and $\kappa$, and therefore,
gives rise to the well-known Coulomb and exchange integrals.

Following the same steps as for the one-particle operators, 
we introduce $C^{(n)}_{\lambda\kappa,\alpha\beta}$ 
in analogy to $C^{(n)}_{\lambda,\alpha}$ of Eq.~(\ref{eq:cofactor}) as
\begin{equation}
 C^{(n)}_{\lambda\kappa,\alpha\beta} = (\pm 1)^{\alpha+\beta+\lambda+\kappa}
 \left\langle \prod_{\mu\neq\lambda,\kappa}^{N_{\text{p}}^{(n)}}
 \phi_{n\mu}(\mathbf{r}_{n\mu})
 \Bigg| \Phi_{n}^\pm(\lambda\kappa|\alpha\beta)
 \right\rangle_{\forall \mathbf{r}_{n\mu\neq n\lambda,n\kappa}} ~.
\end{equation}
The matrix element $O^{[2]}_{nm,\lambda\kappa}$ 
for a two-body operator coupling particles $\lambda,\kappa$
of different particle types $n,m$ is
\begin{align}
 O^{[2]}_{nm,\lambda\kappa}
 &= \left\langle \psi \Big| o_{nm}(\lambda,\kappa) \Big| \psi \right\rangle
 \nonumber \\
 &= \left\langle \Phi_n \Phi_m \Big| o_{nm}(\lambda,\kappa) \Big|
 \Phi_n \Phi_m \right\rangle \prod_{i\neq n,m}^{N_{\text{t}}} \left\langle
 \Phi_i \Big|\Phi_i \right\rangle ~,
\end{align}
with
\begin{align}
 &\left\langle \Phi_n \Phi_m \Big| o_{nm}(\lambda,\kappa) \Big|
 \Phi_n \Phi_m \right\rangle  
 \nonumber \\
 &= \sum_{\alpha}^{N_{\text{p}} ^{(n)}} 
 \sum_{\beta}^{N_{\text{p}} ^{(m)}} 
 C^{(n)}_{\lambda,\alpha} C^{(m)}_{\kappa,\beta}
 \Big{\langle} 
 \phi_{n\lambda}(\mathbf{r}_{n\lambda})
 \phi_{m\kappa}(\mathbf{r}_{m\kappa}) 
 \Big| o_{nm}(\lambda,\kappa) \Big|
 \phi_{n\alpha}(\mathbf{r}_{n\lambda})  
 \phi_{m\beta}(\mathbf{r}_{m\kappa})  
 \Big{\rangle}_{\mathbf{r}_{n\lambda,m\kappa}}
  ~.
\end{align}
Conversely, for operators coupling particles of the same type, it is
\begin{align}
 & O^{[2]}_{n,\lambda\kappa} =
 \prod_{i\neq n}^{N_{\text{t}}} 
 \left\langle \Phi_i \Big| \Phi_i \right\rangle
 \left\langle \Phi_n \Big| o_{n}(\lambda,\kappa) \Big| \Phi_n \right\rangle
 = N_n \left\langle \Phi_n \Big| o_{n}(\lambda,\kappa) \Big| 
 \Phi_n \right\rangle 
 \nonumber \\
 &= N_n \sum_{\alpha<\beta}^{N_{\text{p}} ^{(n)}}
  C^{(n)}_{\lambda\kappa,\alpha\beta}
 \Big{\langle} 
 \phi_{n\lambda}(\mathbf{r}_{n\lambda}) 
 \phi_{n\kappa}(\mathbf{r}_{n\kappa})
 \Big| o_{n}(\lambda,\kappa) (1 \pm \mathcal{P}_{\lambda\kappa}) \Big|
 \phi_{n\alpha}(\mathbf{r}_{n\lambda})  
 \phi_{n\beta}(\mathbf{r}_{n\kappa}) 
 \Big{\rangle}_{\mathbf{r}_{n\lambda,n\kappa}}
 ~.
\end{align}

\subsection{Multi-Stage Variational Optimization} 
\label{subsec:VarOpti}

From Sec.~\ref{subsec:wf} it follows that the trial wave function $\Psi$, 
and hence the energy expectation value, is a function of the linear
combination coefficients $\mathbf{c}=\{c_I\}$, the expansion coefficients $\mathbf{d}=\{d_{Ii\mu\nu}$\}, 
the widths $\bm{\alpha} = \{\alpha_{Ii\mu\nu}\}$ and displacements $\mathbf{s}=\{\mathbf{s}_{Ii\mu\nu}\}$ 
of every Gaussian,
\begin{equation}
  \Psi \equiv \Psi (\mathbf{r};\mathbf{c},\mathbf{d},\bm{\alpha},\mathbf{s}) ~.
  \label{eq:WfnDependence}
\end{equation}
Therefore, approximating bound states of molecular systems translates into determining 
the optimal set of parameters that minimizes the energy.
For a given set of $\bm{\alpha}$, $\mathbf{d}$ and $\mathbf{s}$ parameters,
the Hamiltonian matrix elements can be evaluated analytically and 
the linear coefficients $\mathbf{c}$ are obtained exactly as eigenvectors of $\mathcal{H}_\text{TF}$ represented in the basis. 
By contrast, the optimal set of parameters $(\bm{\alpha},\mathbf{d},\mathbf{s})$ 
must be determined iteratively by variational optimization. 

We construct the trial wave function $\Psi$ bottom-up 
by incorporating and optimizing one basis function $\psi_I$ at a time in order to minimize 
the dimension of the multireference expansion $N_{\rm{b}}$. 
Each time that a new basis function is generated, 
a large pool of randomly chosen Gaussian widths and shifts are separately tested
and only the configuration corresponding to the minimal energy is permanently added to the basis set.
This scheme has been called "competitive selection" and "neighborhood search" 
in Ref.~\citenum{SuVaBook98}.
Since the total energy is a non-linear function of 
the parameters $(\bm{\alpha},\mathbf{d},\mathbf{s})$ with multiple local minima, 
we rely on gradient-free iterative steps, such as the Powell algorithm, to further refine the previously selected parameters 
\cite{Kukulin_1977,Szalewicz_1986,Szalewicz_1987,Szalewicz_1988,SuVaBook98} in a fixed basis set.

The simultaneous variational optimization of the parameters 
$(\bm{\alpha},\mathbf{d},\mathbf{s})$ is equivalent to three steps 
that are usually accomplished separately in calculations based on the BO approximation.
(i) Selecting the most appropriate Gaussian widths $\{\bm{\alpha}\}$ is equivalent to the optimization 
of the atomic orbital basis which is kept fixed in molecular calculations.
(ii) The optimization of the nuclear shifts $\{\mathbf{s}\}$ resembles the geometry optimization 
step. However we stress that in the present work all nuclear-electronic shifts 
are optimized separately and the Gaussians are in general not centered at points 
where the nuclear probability density exhibits a maximum.
Hence, the Gaussian orbitals cannot be regarded as atomic orbitals.
(iii) The optimization of the expansion coefficients $\{\mathbf{d}\}$ corresponds to the
SCF optimization of the molecular orbitals that is carried out with HF or DFT.
This step, different from the previous ones, does not require new integral evaluations, 
which are computationally very demanding for our non-orthogonal orbital basis.
The step can be efficiently carried out by stochastic iterative optimization \cite{Varga1995}.

We note that other multicomponent approaches \cite{Hammes-Schiffer2002,nakai2007nuclear} 
consider a fixed set $\{\bm{\alpha}\}$ that is pre-optimized for each atom. 
On the one hand, this reduces the number of variational parameters to optimize and reduces the
complexity of the delicate non-linear optimization problem.
On the other hand, it reduces the flexibility of the basis and 
many more functions are needed to achieve a given accuracy.

Orbitals $\phi_{Ii\mu}$ in Eq.~(\ref{eq:conf}) defining a configuration 
$\psi_I$ are optimized separately, and hence, orbitals contained in different basis 
functions (in general) do not span the same space. 
This is a decisive difference to HF-based methods 
in which two different configurations might differ by few "excitations" of the one-particle functions. 
While the HF equations achieve the optimal set of orbitals minimizing 
the single-configurational energy, 
we highlight that the stochastic optimization yields the optimal orbitals for which the energy of the 
multireference wave function is minimal.

\section{The Multicomponent Fock Space for the DMRG Algorithm}
\label{sec:DMRGTheory}

We can further improve our solution of the time-independent Schr\"odinger equation by forming a multicomponent full-CI wave function with the NEAP wave function as the starting point.
We first define an overall orbital basis including all single-particle functions of the different, previously optimized $N_b$ NEAP configurations.
We then search for the lowest-energy wave function in the space of all possible configurations constructed in this overall orbital space, \textit{i.e.} the multicomponent full-CI wave function.
However, the size of this full-CI expansion grows factorially with the size of the orbital basis, that in turns grows linearly with $N_b$.
This makes the diagonalization of the Hamiltonian in the full-CI basis prohibitive already for small $N_b$. 
As we will discuss in this section, we overcome this limitation by employing the DMRG algorithm.

\subsection{The Multicomponent Full-CI Space}
\label{subsec:MultiComponentFullCI}

The DMRG algorithm is a powerful technique to diagonalize Hamiltonians in a full-CI basis. 
It can be applied to any wave function that is expressed as a product of one-particle basis functions --
such as the generalized CI expansion $\Psi$ in Eq.~(\ref{eq:GeneralizedCI}) -- subsequent to the definition 
of the multicomponent Fock space.
First, we consider the set $\mathcal{B}_{i}$ as the union of $L_i$ linearly independent 
molecular orbitals for the particles of type $i$ as
\begin{equation}
 \mathcal{B}_{i} = \bigcup_{I=1}^{N_{\text{b}}} \{ \varphi_{Ii\nu}: \nu\in[1,\dots,N_\text{p}^{(i)}] \} ~,
 \label{eq:singlePtypeBasis}
\end{equation}
with
\begin{align}
  L_i\equiv{\text{dim}}(\mathcal{B}_i) = N_{\text{b}} \times N_{\text{p}}^{(i)} \, ,
  \label{eq:DimensionSingleParticleBasis}
\end{align}
and $N_{\text{p}}^{(i)}$ being the total number of particles of type $i$.
Then, we obtain the set $\mathcal{B}$ that contains the molecular orbitals of all particle types as
\begin{equation}
 \mathcal{B} = \bigcup_{i=1}^{N_{\text{t}}} \mathcal{B}_i ~,
 \label{eq:AllPtypeBasis}
\end{equation}
with dimension
\begin{align}
  L\equiv{\text{dim}}(\mathcal{B}) = \sum_i^{N_\text{t}} L_i = N_b \sum_i^{N_\text{t}} N_{\text{p}}^{(i)} \, .
  \label{eq:DimensionAllParticleBasis}
\end{align}
Given the one-particle vector space $\mathcal{V}_1^{(i)}$ generated by the set $\mathcal{B}_i$ , i.e. $\mathcal{V}_1^{(i)} = {\text{span}}(\mathcal{B}_{i})$, we can define the space spanned by all Slater determinants (permanents) for $N^{(i)}_{\text{p}}$ particles of type $i$ as 
\begin{equation}
  \mathcal{V}_{N_{\text{p}}^{(i)}}^{\pm} = \mathcal{S}^{(i)}_{\pm} \mathcal{V}_{N_{\text{p}}^{(i)}}
  = \mathcal{S}^{(i)}_{\pm} \big( \underset{ N_{\text{p}}^{(i)} {\text{ times}} }
    { \underbrace{ \mathcal{V}_1^{(i)} \otimes \mathcal{V}_1^{(i)} \cdots \otimes \mathcal{V}_1^{(i)} }} \big) ~,
  \label{eq:DimensionManyBodySpace}
\end{equation}
where $\mathcal{S}^{(i)}_{\pm}$ (anti)symmetrizes the tensor product of the single particle spaces. 
The resulting space $\mathcal{V}_{N_{\text{p}}^{(i)}}^{\pm}$ is a subset of the full, 
many-body space $\mathcal{V}_{N_{\text{p}}^{(i)}}$ with dimension  
\begin{equation}
  \dim \left( \mathcal{V}_{N_{\text{p}}^{(i)}}^{\pm} \right) = \left( 
    \begin{array}{c} 
      2L_i \\ 
      N_{\text{p}}^{(i)}
    \end{array} \right)
  \label{eq:DimensionFermionic}
\end{equation}
for fermions, and
\begin{equation}
  \dim \big( \mathcal{V}_{N_{\text{p}}^{(i)}}^{\pm} \big) = L_i^{N_{\text{p}}^{(i)}} ~
  \label{eq:DimensionBosonic}
\end{equation}
for bosons. We recall that we construct the $\mathcal{V}_{N_{\text{p}}^{(i)}}$ set from non-orthogonal randomly generated molecular orbitals following the algorithm sketched above. The full-CI space for an ensemble of different particles, e.g. electrons and nuclei, is the direct product of the $\mathcal{V}_{N_{\text{p}}^{(i)}}^{\pm}$ space for each particle type, i.e.
\begin{equation}
  \mathcal{V}^{\text{MC}} = 
  \bigotimes_{i=1}^{N_{\text{t}}} \mathcal{V}_{N_{\text{p}}^{(i)}}^{\pm} ~,
  \label{eq:MultiComponentFockSpace}
\end{equation}
The wave function $\Psi$ given in Eq.~(\ref{eq:GeneralizedCI}) is a product of elements of the single particle-type spaces $\mathcal{V}_{N_{\text{p}}^{(i)}}^{\pm}$ and is, therefore, an element of the vector space $\mathcal{V}^{\text{MC}}$.

\subsection{The Multicomponent Fock Space}

We now define the Fock space for a multicomponent system. We highlight that our derivation follows the one of Ref.~\citenum{streltsov2010general} for the Multi Configurational Time Dependent Hartree (MCTDH), and is closely related to $n$-mode second quantization employed in vibrational-structure calculations \cite{Christiansen2004_SecondQuantization,Wang2009_SQMCTDH}. In the previous section, we defined the vector space $\mathcal{V}_{m^{(i)}}$ for a fixed number $m^{(i)}$ of particles of type $i$. The Fock space $\mathcal{F}_i$ for particles of type $i$ is the direct sum of the spaces $\mathcal{V}_{m^{(i)}}$ for all possible values of $m^{(i)}$.
\begin{align}
  \mathcal{F}_i = \bigoplus_{m^{(i)}=0}^{+\infty} \mathcal{S}^{(i)}_{\pm} \mathcal{V}_{m^{(i)}} \, ,
  \label{eq:FockSpace}
\end{align}
where, for the sake of simplicity, states with up to $N_{\text{p}}^{(i)}$ particles of type $i$ ($m^{(i)}=0, \ldots, N_{\text{p}}^{(i)})$ are considered in the following (in the exact definition of the Fock space, no upper bound is set). The elements of $\mathcal{F}_i$ are occupation number vectors $ \ket{ \sigma^{(i)}_{1}, \cdots, \sigma^{(i)}_{L_i} }$, where $\sigma^{(i)}_\mu$ is the occupation number related to the single-particle orbital $\mu \equiv{\left( I \nu \right)}$ for the particle set $i$. For a spin-orbital basis of spin-$\frac{1}{2}$ fermions (e.g., electrons) each single-particle basis function can be either occupied or unoccupied, i.e. $|\sigma^{(i)}_\mu\rangle \in \{|0\rangle,|1\rangle\}$. In an orbital basis, however, there will be four possibilities, i.e. the orbital can be either empty, occupied with a spin up or spin down particle, or doubly occupied, $\ket{ \sigma^{(i)}_\mu } \in \{ \ket{0}, \ket{\uparrow}, \ket{\downarrow}, \ket{\uparrow\downarrow} \}$. 
By contrast, a single-particle basis function of a bosonic spin-0 particle of a 
given type can be occupied by an arbitrary number of particles of that type.

The multicomponent Fock space $\mathcal{F}^{\text{MC}}$ is obtained from the direct product of the Fock spaces of each particle type and is, therefore, defined as:
\begin{equation}
  \mathcal{F}^{\text{MC}} = \bigotimes_{i=1}^{N_{\text{t}}} \bigoplus_{m^{(i)}=0}^{N_{\text{p}}^{(i)}} 
						    \mathcal{S}^{(i)}_{\pm} \mathcal{V}_{m^{(i)}} \, .
  \label{eq:MulticomponentFock}
\end{equation}
Consequently, a state $\ket{\bm{\sigma}}$ in the Fock space $\mathcal{F}^{\text{MC}}$ is a direct product of occupation number vectors (ONVs), one for each particle type, 
\begin{equation}
  \ket{\bm{\sigma}} = \ket{ \bm{\sigma}_1} \otimes \ket{\bm{\sigma}_2} \otimes
                      \cdots \otimes \ket{ \bm{\sigma}_{N_{\text{t}}}} 
                    = \ket{ \bm{\sigma}_1, \bm{\sigma}_2, \cdots, \bm{\sigma}_{N_{\text{t}}}} \,
  \label{eq:Multicomponent_ONV}
\end{equation}
with, 
\begin{equation}
  \ket{ \bm{\sigma}_i } = \ket{ \sigma^{(i)}_1} \otimes \ket{\sigma^{(i)}_2} \otimes
                      \cdots \otimes \ket{ \sigma^{(i)}_{L_i}} = \ket{ \sigma^{(i)}_{1}, \sigma^{(i)}_{2},\cdots, \sigma^{(i)}_{L_i} }  \, .
  \label{eq:SingleComponentONV}
\end{equation}
We can then introduce the second-quantization equivalent of the generalized CI expansion given in Eq.~(\ref{eq:GeneralizedCI}) as
\begin{equation}
  \ket{\Psi}  =  \sum_{ {\bm{\sigma}}_1 \bm{\sigma}_2 \cdots {\bm{\sigma}}_{N_{\text{t}}}}^{N_\mathrm{FCI}}
                 c_{\bm{\sigma}_1 \bm{\sigma}_2 \cdots \bm{\sigma}_{N_{\text{t}}} } 
                 \ket{ \bm{\sigma}_1} \otimes \ket{\bm{\sigma}_2} \otimes\cdots\otimes 
                 \ket{ \bm{\sigma}_{N_{\text{t}}}}
              =  \sum_{\bm{\sigma}} c_{\bm{\sigma}} |\bm{\sigma}\rangle \, .
 \label{eq:GeneralizedCI-SQ}
\end{equation}
where $c_{ {\bm{\sigma}}_1{\bm{\sigma}}_2\dots{\bm{\sigma}}_{N_{\text{t}}} }$ is the so-called CI coefficient tensor that is obtained by diagonalizing the Hamiltonian. 
We note that the generalized CI expansion given in Eq.~(\ref{eq:GeneralizedCI}) includes only a subset of all possible ONVs for a given single-particle basis. 
For this reason, Eq.~(\ref{eq:GeneralizedCI}) can be considered as a truncation of the full-CI expansion of Eq.~(\ref{eq:GeneralizedCI-SQ}). 

We now determine the size of the multicomponent full-CI space with respect to $N_b$ and the number of particles for each type. The size of the spin-$\frac{1}{2}$ fermion Fock subspace of particle type $i$ in a spin-orbital basis is
\begin{equation}
 N^{f(i)}_\mathrm{FCI} = 
   \begin{pmatrix} 2L^f_i \\ N^{f(i)}_{\text{p}} \end{pmatrix} ~,
\end{equation}
and the one of spin-0 boson particle $j$ is \cite{streltsov2010general}
\begin{equation}
 N^{b(j)}_\mathrm{FCI} = 
 \begin{pmatrix} 
   N^{b(j)}_{\text{p}} + L^b_j -1 \\ 
   N^{b(j)}_{\text{p}} 
 \end{pmatrix} ~.
\end{equation}
Here, $L^f_i$ and $L^b_j$ are the dimensions of the subspaces of particles $i$ and $j$ which are of fermionic '$f$' and bosonic '$b$' nature,
respectively. The overall number of possible configurations is, therefore,
\begin{equation}
 N_\mathrm{FCI} = \prod_i^{N_{\text{t}}} N^{(i)}_\mathrm{FCI} ~,
 \label{eq:OverallNumber}
\end{equation}
where $N_\mathrm{FCI}^{(i)}$ is a substitute for either $N_\mathrm{FCI}^{f(i)}$ or $N_\mathrm{FCI}^{b(i)}$ depending on the type of particle $i$. Hence, we optimize a wave function in a basis of many-body basis functions which yields optimized one-particle functions (orbitals). Subsequently, these orbitals are used to construct a new basis in the mixed boson-fermion Fock space to diagonalize the Hamiltonian.

\subsection{Full Molecular Hamiltonian in Second Quantization}

To encode operators in the ONV space, it is convenient to introduce the elementary second-quantization operators for the multicomponent Fock space. For a fermionic particle of type $i$, we introduce a pair of creation and annihilation operators, $a_{i \mu s}^\dagger$ and $a_{i \mu s}$, that create and annihilate a particle in the basis function $\mu$ with spin $s$, respectively. The fermionic anticommutation rules for these operators read
\begin{equation}
 \begin{aligned}
   \{a^\dagger_{i \mu s}, a^\dagger_{i \nu s^\prime}\} &= 0 \\
   \{a_{i \mu s}, a_{i \nu s^\prime} \} &= 0 \\
   \{a^\dagger_{i \mu s}, a_{i \nu s^\prime}\} &= \delta_{\mu\nu}\delta_{ s s^\prime} ~,
 \end{aligned}
 \label{eq:CommutationRulesFermions}
\end{equation}
where $\{\cdot , \cdot\}$ denotes an anticommutator. Similarly, we introduce a pair of creation and annihilation operators, $b_{i \mu s}^\dagger$ and $b_{i \mu s}$, respectively, for a bosonic particle of type $i$ which fulfill the following commutation relations 
\begin{equation}
 \begin{aligned}
  [b^\dagger_{i \mu s}, b^\dagger_{i \nu s^\prime}] &= 0 \\
  [b_{i \mu s}, b_{i \nu s^\prime}] &= 0 \\
  [b^\dagger_{i \mu s}, b_{i \nu s^\prime}] &= \delta_{\mu\nu}\delta_{ss^\prime} ~,
 \end{aligned}
 \label{eq:CommutationRulesBosons}
\end{equation}
where $[\cdot , \cdot]$ abbreviates a commutator. Even if Eqs.~(\ref{eq:CommutationRulesFermions}) and (\ref{eq:CommutationRulesBosons}) are general, in the present work we will consider only spin-$\frac{1}{2}$ fermions (e.g., electrons and protons) and spin-0 bosons (e.g., He$^4$). We also emphasize that two operators belonging to different particle types always commute since they act on different Hilbert spaces. The full molecular Hamiltonian 
can then be expressed in the second-quantization format following the same procedure as in the electronic structure case. The only difference is that the nucleus-electron attraction and the nucleus-nucleus repulsion are two-body operators, while in the clamped-nuclei picture of the BO approximation they are one-body operators and constants, respectively. Therefore, the second-quantization form of Eq.~(\ref{eq:GeneralizedCI-SQ}) is given by

\begin{equation}
    \begin{aligned}
        \mathcal{H} &= \sum_{i}^{N^f_t} \sum_{\substack{\mu\nu\\ s}}^{L^f_i}  
              t^{(i)}_{\mu\nu}~ a^\dagger_{i\mu s}a_{i\nu s}
           + \sum_{i}^{N^b_t} \sum_{\mu\nu}^{L^b_i}  
              t^{(i)}_{\mu\nu}~ b^{\dagger}_{i\mu} b_{i\nu}
           + \frac{1}{2} \sum_{ij}^{N^f_t} \sum_{\substack{\mu\nu\\ s}}^{L^{f}_i}
             \sum_{\substack{\kappa\lambda \\ s^\prime}}^{L^{f}_j}
              \left( V^{(ij)}_{\mu\nu\kappa\lambda} + T^{(ij)}_{\mu\nu\kappa\lambda} \right)~ 
              a^\dagger_{i\mu s} a^\dagger_{j\nu s^\prime}   
              a_{j\lambda s^\prime} a_{i\kappa s} \\
          &+ \frac{1}{2} \sum_{ij}^{N^b_t} \sum_{\mu\nu}^{L^b_i} \sum_{\kappa\lambda}^{L^b_j}
             \left( V^{(ij)}_{\mu\nu\kappa\lambda} + T^{(ij)}_{\mu\nu\kappa\lambda} \right)~ 
             b^\dagger_{i\mu} b^\dagger_{j\nu} b_{j\lambda} b_{i\kappa}
           + \sum_{i}^{N^f_t} \sum_{j}^{N^b_t}  \sum_{\substack{\mu\nu \\ s}}^{L^{f}_i}
             \sum_{\kappa\lambda}^{L^b_j}  
             \left( V^{(ij)}_{\mu\nu\kappa\lambda} + T^{(ij)}_{\mu\nu\kappa\lambda} \right)~ 
             a^\dagger_{i\mu s} b^\dagger_{j\kappa} b_{j\lambda} a_{i\nu s} ~.
    \end{aligned}
    \label{eq:SQPreBO}
\end{equation}
The integrals $t_{\mu\nu}^{(i)}$ is the one-body kinetic energy of a particle of type $i$ minus the one-body part of the kinetic energy of the center-of-mass (see Eq.~(\ref{eq:TranslationallyInvariantHam}) \textit{e.g.}, for an orthogonal basis set $\{ \zeta_{i,\mu} \}$

\begin{equation*}
    t_{\mu\nu}^{(i)}
      = \Big{\langle} \zeta_{i\mu} (\mathbf{r}_{i\mu})\Big{|} \left(-\frac{1}{2 m_i}+\frac{1}{2M}\right)  \bm{\nabla}^2_{\bm{r}_{i\mu}} \Big{|} \zeta{i\nu} (\mathbf{r}_{i\mu})\Big{\rangle} \, ,
\end{equation*}

and the integrals $T_{\mu\nu\kappa\lambda}^{(ij)}$ contain the matrix elements of the two-body part of the kinetic energy operator of the center-of-mass (see Eq.~(\ref{eq:TranslationallyInvariantHam})) \textit{e.g.}, for an orthogonal basis set $\{ \zeta_{i,\mu} \}$

\begin{equation*}
    T_{\mu\nu\kappa\lambda}^{(ij)}
    =
    \Big{\langle} 
    \zeta_{i\mu}(\mathbf{r}_{i\mu})
    \zeta_{j\nu}(\mathbf{r}_{j\nu}) 
    \Big|  
    \frac{1}{M}\bm{\nabla}_{\bm{r}_{i\mu}}\bm{\nabla}_{\bm{r}_{j\nu}}
    \Big|
    \zeta_{i\kappa}(\mathbf{r}_{i\mu})  
    \zeta_{j\lambda}(\mathbf{r}_{j\nu})  
    \Big{\rangle}
\end{equation*}

We stress here that these two-body kinetic energy operators emerge from the elimination of the center-of-mass contributions in Laboratory Fixed Cartesian coordinates as derived in Eq.~(\ref{eq:TranslationallyInvariantHam}).
The $V^{(ij)}_{\mu\nu\kappa\lambda}$ parameters are the integrals of the Coulomb interaction between  particles of two types $i$ and $j$ calculated for the orbitals $\phi_{i\mu}$, $\phi_{i\kappa}$ for particle type $i$, and $\phi_{j\nu}$, $\phi_{j\lambda}$ for particle type $j$ ($i$ and $j$ may be identical). 

The full molecular Hamiltonian is given by a many-body expansion that terminates exactly at the second order. This is a major difference compared to the BO approximation, which produces the vibrational Schr\"{o}dinger equation to be solved for a given PES, that is obtained from the solution of the electronic Schr\"{o}dinger equation. A many-body expansion of a PES, for example obtained through its $n$-mode representation \cite{Carter1997_VCI,Kongsted2006_NMode}, does not terminate at any order. As we will discuss in the next paragraph, the absence of high-order couplings in the Hamiltonian makes the application of DMRG to the full molecular Hamiltonian particularly appealing.

\subsection{Matrix Product States and Operators}
\label{subsec:DMRG}

The generalized full CI wave function of Eq.~(\ref{eq:GeneralizedCI-SQ}) can be optimized by minimizing the total
energy obtained as the expectation value of the Hamiltonian in Eq.~(\ref{eq:SQPreBO}).
This leads to the pre-BO full-CI algorithm, whose range of application is, however, limited by the steep scaling of its computational cost with the number of particles and basis functions. 
We have already mentioned that we aim to reduce the computational cost of the CI step with DMRG.
A detailed presentation of the DMRG and of its applications in quantum chemistry can be found in several reviews \cite{schollwock2005density,Marti2008_Review,Chan2008_Review,Legeza2008_Book,Zgid2009_Review,Marti2011_Review,schollwock2011density,chan2011density,Wouters2014_Review,Kurashige2014_Review,szalay2015tensor,Olivares2015_DMRGInPractice,Yanai2015,Baiardi2019_DMRG-Review}. Here, we only present the main features of DMRG to eventually describe its extension to the full molecular Hamiltonian.

DMRG is an optimization algorithm for wave functions $\Psi_{\text{MPS}}$ encoded as matrix product states (MPS),
\begin{equation}
 \ket{\Psi_{\text{MPS}}} = \sum_{\bm{\sigma}} \sum_{a_1,a_2\dots,a_{L-1}}^m
    M_{1,a_1}^{\sigma_1} M_{a_1,a_2}^{\sigma_2} \cdots M_{a_{L-1},1}^{\sigma_L} \ket{ \bm{\sigma} }
 \label{eq:MPS-Parametrization}
\end{equation}
By comparing Eq.~(\ref{eq:MPS-Parametrization}) with Eq.~(\ref{eq:GeneralizedCI-SQ}), we note that an MPS will be equivalent to a CI wave function if the $c_{\bm{\sigma}}$ tensor is expressed as
\begin{equation}
  c_{\sigma_1\sigma_2 \dots \sigma_L} 
    = \sum_{a_1,a_2\dots,a_{L-1}}^{m} M_{1,a_1}^{\sigma_1} M_{a_1,a_2}^{\sigma_2} \cdots M_{a_{L-1},1}^{\sigma_L}
    = \bm{M}^{\sigma_1} \bm{M}^{\sigma_2} \cdots \bm{M}^{\sigma_L} ~.
  \label{eq:TT-CI-Tensor}
\end{equation}
In numerical analysis, Eq.~(\ref{eq:TT-CI-Tensor}) is called the tensor-train (TT) factorization of the CI tensor $c_{\bm{\sigma}}$ \cite{Oseledets2011_TTGeneral}. Intuitively, the TT factorization expresses the $L$-rank CI tensor $c_{\bm{\sigma}}$ as a product of $L$ rank-three tensors $M_{a_{i-1},a_i}^{\sigma_i}$ (note, however, that $M^{\sigma_1}$ and $M^{\sigma_L}$ are row and column vectors, respectively). $\sigma_i$ corresponds to one of the indices of the original tensor $c_{\bm{\sigma}}$, while the two additional indices, $a_{i-1}$ and $a_i$, are auxiliary dimensions that are introduced in the factorization of $c_{\bm{\sigma}}$. The accuracy of the TT factorization depends on the maximum dimension of the $a_i$ indices, a parameter that is usually referred to as \quotes{bond dimension} or \quotes{number of block states} and indicated as $m$. A CI wave function can be encoded exactly as in Eq.~(\ref{eq:TT-CI-Tensor}) with a bond dimension that grows exponentially with the size of the one-particle basis (usually defined \quotes{lattice} in DMRG). 
The area law \cite{Hastings2007_AreaLaw} ensures that, for a given accuracy, 
the ground state of any gapped Hamiltonian with short-range interactions can be encoded
as an MPS with a bond dimension $m$ that is independent of the lattice size. The full molecular Hamiltonian contains terms with 
long-range Coulombic pair interactions producing 4-index parameters in its second-quantization form, 
and therefore, the area law does not apply. Nevertheless, it has been shown that the DMRG is efficient enough for electronic and vibrational Hamiltonians \cite{Marti2008_DMRGMetalComplexes,Boguslawski2012_SpinDensities,Yanai2013_Photosystem,Chan2014_Fe2S2,Yanai2014_Desaturase,Oseledts2016_VDMRG,Baiardi2017_VDMRG} to render full-CI calculations feasible on systems with up to $L$=100. One aim of this work will be to probe whether this holds also true for full molecular Hamiltonians. 

As for the wave function, we encode also the Hamiltonian as a matrix product operator (MPO)
\begin{equation}
  \mathcal{H} = \sum_{\bm{\sigma},\bm{\sigma}^\prime}^{N_\mathrm{FCI}} \sum_{b_1,b_2\dots,b_{L-1}}^{b_{\text{max}}} 
  H_{1,b_1}^{\sigma_1,\sigma_1^\prime} H_{b_1,b_2}^{\sigma_2,\sigma_2^\prime} \cdots H_{b_{L-1},1}^{\sigma_L,\sigma_L^\prime}
  \ket{\bm{\sigma}} \bra{\bm{\sigma}^\prime} ~.
  \label{eq:mpo}
\end{equation}
Unlike an MPS, which approximates a CI wave function for finite values of the bond dimension $m$, the MPO-encoded Hamiltonian in Eq.~(\ref{eq:mpo}) is still exact. This encoding is accomplished by starting from its second-quantization form following the algorithm introduced in Ref.~\citenum{Crosswhite2008_MPO-FiniteAutomata} and applied to electronic \cite{keller2015efficient,keller2016spin} and vibrational \cite{Baiardi2017_VDMRG,Baiardi2019_HighEnergy-vDMRG} Hamiltonians. An MPS can be optimized with the so-called alternating least-squared (ALS) algorithm. The tensors $\bm{M}^{\sigma_i}$ are optimized sequentially, one after the other, starting at the first lattice site and going back and forth along the DMRG chain. The optimization of the individual site is called a micro-iteration step, and the one of the whole chain (back and forth) is called a macro-iteration or \textit{sweep}. Starting from Eqs.~(\ref{eq:MPS-Parametrization}) and (\ref{eq:mpo}), minimization of the energy of site $j$ leads to the following eigenvalue equation \cite{schollwock2011density}
\begin{equation}
  \sum_{\sigma_i^\prime} \sum_{a^\prime_{i-1},a^\prime_i}^{m} \sum_{b_{i-1},b_i}^{b_{\text{max}}} 
    L_{a_{i-1},a^\prime_{i-1}}^{b_{i-1}} ~ H_{b_{i-1},b_i}^{\sigma_i,\sigma_i^\prime} ~
    R_{a_i,a^\prime_i}^{b_i} ~ M_{a^\prime_{i-1},a_i^\prime}^{\sigma^\prime_i} = 
    E \, M_{a_{i-1},a_i}^{\sigma_i} ~,
  \label{eq:DMRG_Eigenvalue}
\end{equation}
where the rank-3 tensors $L_{a_{i-1},a_{i-1}^\prime}^{b_{i-1}}$ and $R_{a_i,a^\prime_i}^{b_j}$, usually defined as left and right boundary, respectively, collect the contraction of the MPS with the MPO for all sites different from the $j$-th one. Eq.~(\ref{eq:DMRG_Eigenvalue}) is solved with iterative solvers, such as the Davidson algorithm.
The DMRG calculations are repeated for increasing values of $m$ to probe convergence with respect to the bond dimension $m$. The overall algorithm is known as single-site DMRG.
Alternatively, the tensors of two consecutive sites can be optimized simultaneously within the so-called two-sites formulation of DMRG that is more stable and less likely to converge to local minima of the energy functional \cite{schollwock2011density}.
If not otherwise stated, we carried out all calculations with the two-site variant of DMRG.

\subsection{Symmetries}

The anticommutation properties of fermionic operators are not included automatically in the MPO. We enforce the proper permutational symmetry by applying the Jordan--Wigner transformation \cite{jordan1993paulische} to transform fermionic operators to their bosonic counterpart. We introduce the fermionic filling operator acting on site $\mu$ as \cite{keller2015efficient}
\begin{equation}
 \mathcal{F}_\mu = 
 \left(
   \begin{array}{c|cccc}
       & |0\rangle & |\uparrow\rangle &  |\downarrow\rangle & |\uparrow\downarrow\rangle \\ 
       \hline
      \ket{0}          & 1 & 0  & 0 & 0 \\
      \ket{\uparrow}   & 0 & -1 & 0 & 0 \\
      \ket{\downarrow} & 0 & 0 & -1 & 0 \\
      \ket{\uparrow\downarrow} & 0 & 0 & 0 & 1 \\
   \end{array} 
 \right)
\end{equation}
The representation of $\mathcal{F}_\mu$ in the single-orbital second-quantization basis is a diagonal matrix with a `$-1$` entry for each state with an odd number of fermions and `$+1$` for even numbers of fermions. The Jordan--Wigner transformation maps the fermionic creation ($a_{\mu\uparrow}^\dagger$) and annihilation ($a_{\mu\uparrow}$) operators for a spin-up fermion to its bosonic counterpart as
\begin{equation}
 \begin{aligned}
  a_{\mu\uparrow}         &\mapsto \mathcal{F}_1 \dots \mathcal{F}_{\mu-1}  b_{\mu\uparrow} \\
  a^\dagger_{\mu\uparrow} &\mapsto \mathcal{F}_1 \cdots \mathcal{F}_{\mu-1} b^\dagger_{\mu\uparrow}    ~.
 \end{aligned} 
 \label{eq:JW-AlphaElectron}
\end{equation}
The definition of the transformation for the creation ($a_{\mu\downarrow}^\dagger$) and annihilation ($a_{\mu\downarrow}$) operators for a spin-down fermion is similar, but an extra filling operator must be added for site $\mu$
\begin{equation}
 \begin{aligned}
  a_{\mu\downarrow} &\mapsto \mathcal{F}_1 
   \dots  \mathcal{F}_{\mu-1}  \mathcal{F}_\mu b_{\mu\downarrow} \\  
  a_{\mu\downarrow}^\dagger &\mapsto \mathcal{F}_1 \dots \mathcal{F}_{\mu-1}  b^\dagger_{\mu\downarrow} 
  \mathcal{F}_\mu   ~. 
 \end{aligned}
 \label{eq:JW-BetaElectron}
\end{equation}
We note that creation and annihilation operators do not commute with the filling operator when acting on the same site. From Eqs.~(\ref{eq:JW-AlphaElectron}) and (\ref{eq:JW-BetaElectron}) we can derive the Jordan-Wigner form of a one-particle term coupling the orbitals $\mu$ and $\nu$ as
\begin{equation}
 \begin{aligned}
  a^\dagger_{\mu\uparrow} a_{\nu\uparrow}  &\mapsto  a^\dagger_{\mu\uparrow} 
 	 \mathcal{F}_{\mu}  \mathcal{F}_{\mu+1} \cdots \mathcal{F}_{\nu-1} a_{\nu\uparrow}   \\ 
  a^\dagger_{\mu\downarrow} a_{\nu\downarrow}  &\mapsto   a^\dagger_{\mu\downarrow}  
	 \mathcal{F}_{\mu+1} \cdots \mathcal{F}_{\nu-1} \mathcal{F}_{\nu} a_{\nu\downarrow} ~. 
 \end{aligned}
 \label{eq:JW-OneBodyTerm}
\end{equation}
Eq.~(\ref{eq:JW-OneBodyTerm}) can be easily extended to two-particle operators. As a rule, the Jordan--Wigner mapping transforms a one-body operator that couples orbitals $\mu$ and $\nu$ by adding a filling operator for all orbitals $\kappa$ for which $\mu<\kappa<\nu$. It is, therefore, clear that different representations are obtained depending on the sorting of the orbitals in the DMRG lattice. This is, however, not surprising because any second-quantization representation depends on the sorting of the orbitals as well. In DMRG, we first apply the Jordan--Wigner transformation and then encode the resulting operator as an MPO, in analogy to what is done for non-relativistic electronic-structure theory \cite{Dolfi2014_ALPSProject,keller2015efficient}. Applying the Jordan--Wigner transformation to the MPO form of the full molecular Hamiltonian poses, however, an additional challenge. The DMRG lattice, i.e. the set of all local basis functions, includes the orbitals associated with all possible particle types. The Jordan--Wigner transformation must be applied to the second-quantization operators of all fermionic particles, which can be either the electrons or fermionic nuclei (such as a proton). Therefore, the transformation rules of Eqs.~(\ref{eq:JW-AlphaElectron}) and (\ref{eq:JW-BetaElectron}) apply separately to each particle type. To map a one-particle term such as the one given in Eq.~(\ref{eq:JW-OneBodyTerm}) for particle type $i$, the filling operator must be applied only to the sites that are associated with particle type $i$.

The presence of terms with equal numbers of creation and annihilation operators for each particle type implies that the Hamiltonian in Eq.~(\ref{eq:SQPreBO}) conserves the particle number of each type. 
Formally, it is invariant upon unitary transformations between the orbitals of a given particle type.
We follow the procedure discussed in Refs.~\citenum{Vidal2011_DMRG-U1Symm,Troyer2011_PEPS-Symmetry} to reduce the computational effort in contracting the $\bm{M}^{\sigma}$ tensors with an MPO based on this symmetry.
The computational cost of NEAP-DMRG could be further reduced by exploiting the conservation of the total spin symmetry. For the non-relativistic electronic Hamiltonian, this implies that the wave function is invariant under transformations of the SU(2) non-Abelian group. Spin-adapted formulations of DMRG \cite{McCulloch2002_NonAbelianDMRG,Zgid2008_SpinAdaptation,Wouters2012,Sharma2015_GeneralNonAbelian,keller2016spin} can be extended to any particle type, but this is beyond the scope of this work.

\subsection{L{\"o}wdin Orthonormalization}
\label{APP:Lowdin}

Non-orthogonal sets of electron-nuclear single-particle functions increase the flexibility of each trial multi-component configuration, and hence, allow us to obtain a compact NEAP wave function. 
Therefore, as we have already noted in Sec.~III A, the NEAP method yields the optimal orbitals for which the energy of the multireference wave function is minimal and the generalized CI wave function is most compact.
Through the DMRG algorithm we further extend this CI-like wave function by combining orbitals of different NEAP configurations. Orthogonalization of the orbital basis does not change the DMRG converged energies because both the non-orthogonal and orthogonalized $\mathcal{B}_i$ sets span the same space. Yet, it is necessary because the standard DMRG algorithm of Section~IV~D requires orthogonal orbital bases
(although that restriction may be lifted).
Also, amid of the orthogonalization, we screen the orbital basis and avoid addition in the MPO of redundant overlapping orbitals.
Near linear dependence of the basis functions in $\mathcal{B}_i$ is identified based on the magnitude of eigenvalues of the overlap matrix $\bm{S}$. Eigenvectors associated to eigenvalues smaller than a given threshold are discarded to reduce systematically the numerical error associated to the orthogonalization step. 
To avoid discarding many orbitals, which would drastically reduce the size of the orthogonal space, we implemented this step in arbitrary precision arithmetic.
In this way, one still has the possibility to reduce the orthogonal space discarding redundant orbitals, while not being affected by any numerical error in the orthogonalization.

For the sake of completeness, we briefly review the essential equations of this procedure.
We consider a transformation matrix $\bm{X} \in \mathbb{R}^{L \times L}$ that transforms non-orthogonal molecular orbitals 
$\{ \varphi_\mu \}_{\mu=1,\dots,L}$ into an orthonormal basis $\{ \zeta_\mu \}_{\mu=1,\dots,L}$ of the $L$-dimensional space
\begin{equation}
 \bm{\zeta} = \bm{X} \bm{\varphi} ~,
 \mbox{with} ~
 \langle\zeta_\mu|\zeta_\nu\rangle = \delta_{\mu\nu} \quad \forall\zeta_\mu,\zeta_\nu~\in~\bm{\zeta} ~.
\end{equation}
We collect all non-orthogonal molecular orbitals in the vector $\bm{\varphi}$ as
$
 \bm{\varphi}^\mathrm{T} = 
 \left(\varphi_1~~\varphi_2~~\dots~~\varphi_L \right),
$
and all orthonormal molecular orbitals in a the vector $\bm{\zeta}$ as
$
 \bm{\zeta}^\mathrm{T} = 
 \left(\zeta_1~~\zeta_2~~\dots~~\zeta_L \right) ~.
$
For the basis to be orthonormal, the transformation must fulfill the following relation
\begin{align}
 \bm{\zeta}\bm{\zeta}^\mathrm{T} =  
 \bm{X} \bm{\varphi} \bm{\varphi}^\mathrm{T} \bm{X}^\mathrm{T}
 = \bm{X} \bm{S} \bm{X}^\mathrm{T} = \mathbb{I} ~,
\end{align}
where we have defined the overlap matrix elements 
in the molecular orbital basis as 
\begin{equation*}
  S_{\mu\nu} = (\bm{\varphi} \bm{\varphi}^\mathrm{T})_{\mu\nu}  
      = \frac{\langle\varphi_\mu| \varphi_\nu\rangle}
             { \left\lVert \varphi_\mu \right\rVert 
      \left\lVert \varphi_\nu \right\rVert } ~, 
      \hspace{0.5cm} \mu,\nu = 1,2,\dots,L ~.
\end{equation*}
Here, $\left\lVert . \right\rVert$ denotes the $L^2$ norm. 
Consequently, the transformation matrix can be parametrized as
$
 \bm{X} = \bm{B} \bm{S}^{-\frac{1}{2}} ~,
$
where $\bm{B}$ is an arbitrary unitary matrix that can be chosen as unity,
leading to the \textit{symmetric othonormalization}
$
 \bm{\zeta} = \bm{S}^{-\frac{1}{2}} \bm{\varphi} ~.
$
One-body integrals are transformed into 
the orthonormal basis according to
\begin{align}
 t_{\mu\nu}  = \Big{\langle} \zeta_\mu \Big| o(\nu) \Big| 
 \zeta_\nu \Big{\rangle} 
 = \sum_{\kappa\lambda}^L X^\mathrm{T}_{\kappa\mu} 
 \frac{ \Big\langle \varphi_\kappa \Big| o(\lambda) \Big| 
 \varphi_\lambda \Big\rangle }
 { \left\lVert \varphi_\kappa \right\rVert 
 \left\lVert \varphi_\lambda \right\rVert }
 X_{\nu\lambda}
 = \sum_{\kappa\lambda}^{L} S^{-\frac{1}{2}}_{\mu\kappa} 
 \Tilde{t}_{\kappa\lambda} S^{-\frac{1}{2}}_{\nu\lambda}~, 
 \label{eq:1body}
\end{align}
and the transformation of the two-body integrals reads
\begin{equation}
   V_{\mu\nu\kappa\lambda} = \sum_{\alpha\beta\gamma\delta}^{L} 
                             S^{-\frac{1}{2}}_{\mu\alpha} S^{-\frac{1}{2}}_{\nu\beta} 
                             \Tilde{V}_{\alpha\beta\gamma\delta}
                             S^{-\frac{1}{2}}_{\kappa\gamma}  S^{-\frac{1}{2}}_{\lambda\delta}, 
 \label{eq:2body}
\end{equation}
where all indices $\mu,\nu,\kappa,\lambda$ run from 1 to $L$ each.

The implementation of Eqs.~(\ref{eq:1body}) and (\ref{eq:2body}) 
in finite-precision arithmetics yields numerical instabilities since 
the inversion of a matrix is required. 
To avoid numerical instabilities in the inversion of almost-singular matrices, 
the $\bm{S}$ matrix is diagonalized and the eigenvalues (and 
their corresponding eigenvectors) that are lower than a given threshold 
are removed. 

The diagonalized overlap matrix is obtained as
$
 \bm{\Lambda} = \bm{U}^\mathrm{T} \bm{S} \bm{U} = 
 \text{diag}(\lambda_1,\lambda_2,\dots,\lambda_L) ~,
$
where $\bm{U} \in \mathbb{R}^{L \times L}$ is a unitary matrix to obtain 
the diagonal form of $\bm{S}$.
The \textit{canonical orthonormalization} is then achieved by computing
\begin{equation}
 \bm{\zeta} = \bm{\Lambda}^{-\frac{1}{2}}  \bm{U}^\mathrm{T} \bm{\varphi} ~,
\end{equation}
since it can be shown that
$
 \bm{\Lambda}^{-\frac{1}{2}}  \bm{U}^\mathrm{T} = \bm{U}^\mathrm{T} \bm{S}^{-\frac{1}{2}} ~.
$
$\bm{\zeta}$ still contains the linearly dependent orbitals 
with their corresponding eigenvalues on the diagonal of $\bm{\Lambda}$ 
that are approximately zero. 
Those orbitals are simply omitted, leading to a new eigenvalue problem
\begin{equation}
 \bm{S} \bm{V} =  \bm{V} \bm{\Tilde{\Lambda}} ~,
\end{equation}
with $\bm{V} \in \mathbb{R}^{L\times M}$, and 
$
 \bm{\Tilde{\Lambda}} = \bm{V}^\mathrm{T} \bm{S} \bm{V} = 
 \text{diag}(\lambda_1, \lambda_2,\dots,\lambda_M), 
$
with $M\le L$.
Therefore, the rectangular matrix $\bm{V} \in \mathbb{R}^{L\times M}$ contains 
the eigenvectors belonging to the eigenvalues $\lambda$ that are not omitted. 
Now, we can write the new transformation as
\begin{equation}
 \bm{\zeta} = \bm{\Tilde{\Lambda}}^{-\frac{1}{2}}  \bm{V}^\mathrm{T} \bm{\varphi} ~,
\end{equation}
so that
%
$
 \bm{X} = \bm{\Tilde{\Lambda}}^{-\frac{1}{2}} \bm{V}^\mathrm{T} ~.
$
%
The integrals are transformed into the new basis according to
\begin{equation}
 t_{\mu\nu} = \sum_{\kappa\lambda}^{L} X^{\mathrm{T}}_{\kappa\mu}
 \Tilde{t}_{\kappa\lambda} X_{\nu\lambda}, \quad \mu,\nu = 1,2,\dots,M ~,
\end{equation}
and
\begin{equation}
  V_{\mu\nu\kappa\lambda} = \sum_{\alpha\beta\gamma\delta}^{L}
  X^{T}_{\alpha\mu} 
  X^{T}_{\beta\nu}
  \Tilde{V}_{\alpha\beta\gamma\delta}
  X_{\kappa\gamma}
  X_{\lambda\delta}, \quad \mu,\nu,\kappa,\lambda = 1,2,\dots,M ~.
  \label{eq:TwoBodyIntegralTransformation}
\end{equation}

The computational cost of the transformation defined in Eq.~(\ref{eq:TwoBodyIntegralTransformation}) scales as $\mathcal{O}(L^5)$, and has to be carried out only once at the end of a NEAP calculation. 
Conversely, a single iteration of the iterative diagonalization of Eq.~(\ref{eq:DMRG_Eigenvalue}) scales as $\mathcal{O}(m^3)$ and must be repeated for each site and for each sweep with an overall cost scaling as $\mathcal{O}(m^3 \times L \times N_\text{sweep})$.
This scaling is, however, only formal since the area law (see Section~\ref{subsec:DMRG}) does not apply to the pre-Born--Oppenheimer Hamiltonian defined in Eq.~(\ref{eq:TranslationallyInvariantHam}), and therefore $m$ changes with $L$. 
For this reason, the transformations given in Eq.~(\ref{eq:TwoBodyIntegralTransformation}) may be in principle the bottleneck for small $m$ values. 
However, for any molecular Hamiltonian, the required bond dimension will be large enough to make the DMRG optimization the bottleneck

\section{Numerical Results}
\label{sec:results}

In this section, we present results obtained for the total ground state energy of three molecular systems, H$_2$, H$_3^+$, and BH$_3$, with the NEAP and NEAP-DMRG methods. 
We implemented the multicomponent DMRG algorithm in the QCMaquis-V software package \cite{Baiardi2017_VDMRG}, which we will make available open source in the near future.
The nucleus-electron mass ratio for the proton is $1836.152701$ \cite{hilico2000ab}, and for the boron nucleus, $^{11}$B, a value 
of $20063.7360451$ \cite{wang2012ame2012} was chosen.

All DMRG calculations were performed with the two-site formulation.
We assembled the DMRG lattice by mapping orbitals of the same particle type to adjacent sites, sorting particle blocks according to increasing particle mass and sorting the orbitals in the same order in which they were added to the NEAP wave function.
This is not the usual choice made in an electronic DMRG, where molecular orbitals are obtained with a SCF procedure and sorted according to the eigenvalue of the Fock operator.

\begin{figure}[htbp!]
  \centering
  \includegraphics[width = 0.75\linewidth]{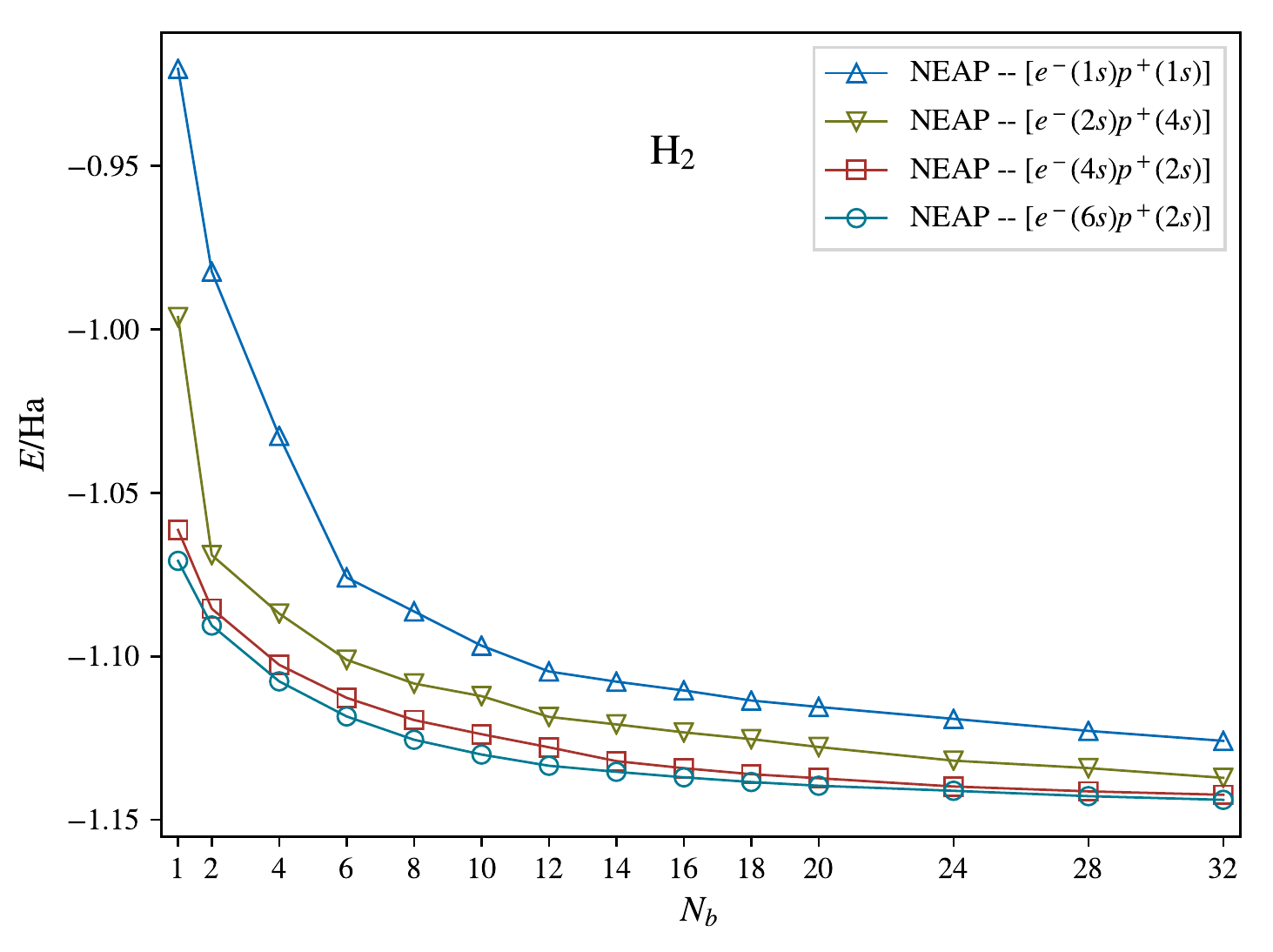}
  \caption{Convergence of the NEAP total ground-state energy of H$_2=\{2$e$^-, 2$p$^+ \}$ with the number of basis functions $N_b$ obtained with $N^{\text{e}^-}_\text{LCPO}=1$ and $N^{\text{p}^+}_\text{LCPO} = 1$ (dark blue line), $N^{\text{e}^-}_\text{LCPO}=2$ and $N^{\text{p}^+}_\text{LCPO}=4$ (green line), $N^{\text{e}^-}_\text{LCPO}=4$ and $N^{\text{p}^+}_\text{LCPO}=3$ (red line), and $N^{\text{e}^-}_\text{LCPO}=6$ and $N^{\text{p}^+}_\text{LCPO}=2$ (light blue line).}
  \label{fig:h2moao}
\end{figure}

First, we consider H$_2$ and study the convergence of NEAP as a function of the number of primitive functions for each orbital ($N_\mathrm{LCPO}$) and the number of basis functions ($N_{\text{b}}$) in the multiconfigurational expansion. 
We compare four different sets of $N^{\text{e}^-}_\text{LCPO}$ and $N^{\text{p}^+}_\text{LCPO}$ values, i.e.
$[\text{e}{^-}(1\text{s})\text{p}{^+}(1\text{s})]$, $[\text{e}{^-}(2\text{s})\text{p}{^+}(4\text{s})]$, $[\text{e}{^-}(4\text{s})\text{p}{^+}(2\text{s})]$, and $[\text{e}{^-}(6\text{s})\text{p}{^+}(2\text{s})]$. The shorthand notation $[\text{e}^-(n\,\text{s})\text{p}^+(m\,\text{s})]$ corresponds to $N^{\text{e}^-}_\text{LCPO}=n$ and $N^{\text{p}^+}_\text{LCPO}=m$, with $n,m\in\mathbb{N}_0$.
In all four cases, we study the convergence of the energy with respect to $N_b$, see the results in Fig.~\ref{fig:h2moao}. 
Since we optimized the primitive basis of each configuration separately, the energy converges to the exact
non-relativistic value for any combination of $N^{\text{p}^+}_\text{LCPO}$ and $N^{\text{e}^-}_\text{LCPO}$ values in the limit $N_b \rightarrow +\infty$.
This appears evident in the right sector of Fig.~\ref{fig:h2moao}, while for smaller $N_b$ values, more flexible primitive basis sets will result in a considerably lower energy. Hence, the difference between the energies of the four different basis sets decreases as $N_b$ increases.

For a given $N_b$ value, the NEAP energy decreases as $N^{\text{e}^-}_\text{LCPO}$ and $N^{\text{p}^+}_\text{LCPO}$ increase.
From Fig.~\ref{fig:h2moao} we infer that increasing $N^{\text{e}^-}_\text{LCPO}$ is more beneficial for energy convergence than increasing $N^{\text{p}^+}_\text{LCPO}$. 
For instance, the basis sets $[\text{e}^-(2\text{s})\text{p}^+(4\text{s})]$ and $[\text{e}^-(4\text{s})\text{p}^+(2\text{s})]$ contain equal number of parameters, but the energies obtained with the latter basis set are consistently lower than the ones with the former for all $N_b$ values.
For this reason, in further calculations we considered a smaller number of primitive functions for the nuclear orbitals than for the electronic ones.

\begin{figure}[htbp!]
  \centering
  \includegraphics[width = 0.75\linewidth]{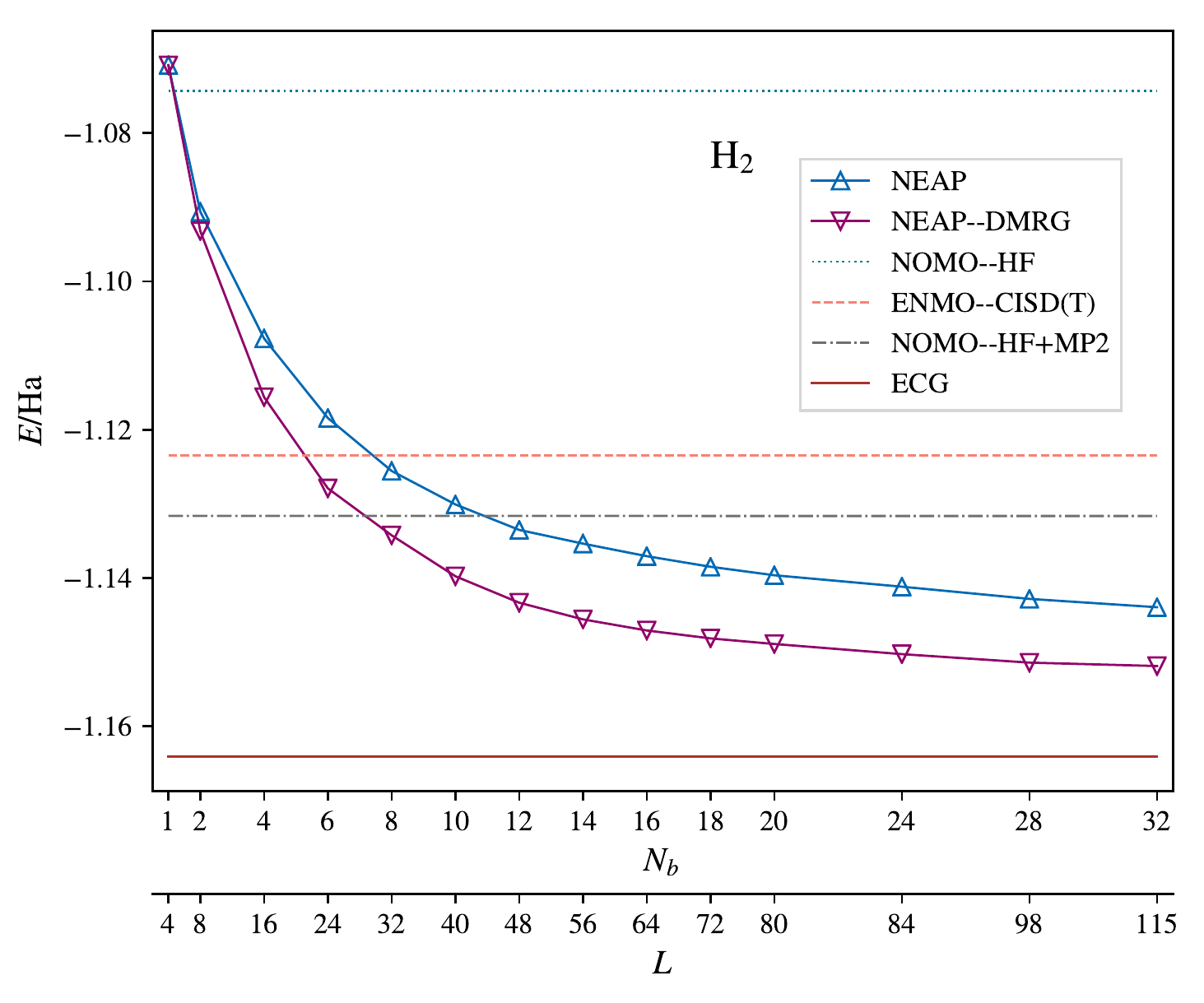}
  \caption{Convergence of NEAP and NEAP-DMRG ($m=60$) energies for the H$_2$ molecule with a $[\text{e}^-(6\text{s})\text{p}^+(2\text{s})]$ basis set with respect to the number of basis functions in the generalized CI expansion ($N_b$) and the dimension of the DMRG lattice ($L$), respectively. The horizontal lines refer to energies obtained with nuclear orbital plus molecular orbital Hartree-Fock (NOMO-HF, blue dotted line)~\cite{nakai2007nuclear}, nuclear orbital plus molecular orbital second-order M{\o}ller-Plesset perturbation theory (NOMO--HF+MP2, black double dotted line)~\cite{nakai2007nuclear}, electron and nuclear molecular-orbital configuration interaction with single, double, and perturbative triple excitations (ENMO--CISD(T), orange dashed line)~\cite{bochevarov2004electron}, and explicitly-correlated Gaussians (ECG, solid red line)~\cite{muolo2018generalized}. The upper bound to the energy is $-1.15226$ Ha (obtained with NEAP-DMRG with $N_b$=32).}
  \label{fig:h2}
\end{figure}

In Fig.~\ref{fig:h2}, we compare the NEAP-DMRG ground-state energies of H$_2$ obtained with the $[\text{e}^-(6\text{s})\text{p}^+(2\text{s})]$ basis with NOMO-HF and NOMO-HF-MP2 results obtained by Nakai and co-workers \cite{nakai2007nuclear}, ENMO-CISD(T) energies by Bochevarov and co-workers \cite{bochevarov2004electron}, and  benchmark results obtained with explicitly correlated Gaussians \cite{muolo2018generalized}. 
The DMRG calculations are converged up to mHa accuracy with a bond dimension $m=60$.
For a given value of $N_b$, the number of DMRG sites $L$ is the dimension of the nuclear and electronic orbital set that is equal to the sum of the number of orbitals for each particle type (see Eq.~\ref{eq:DimensionAllParticleBasis}), e.g. $L=N_b \times 4$ for the H$_2$ molecule, if no basis function is pruned during the L\"owdin orthogonalization.
Note that $N_b$, defined in Eq.~(\ref{eq:GeneralizedCI}), is the number of NEAP basis functions as opposed to the number of full-CI configurations $N_\text{FCI}$ that depends on $L$ as defined in Eqs.~(40)-(42).
From Eq.~(\ref{eq:OverallNumber}) it follows that the configurational space sampled by DMRG is larger than the one sampled by NEAP, and hence, the NEAP-DMRG energies are consistently lower than the NEAP ones for each $N_b$ value. The NEAP-DMRG results with the largest $N_b$ value differ by about~$0.01$~Ha from the explicitly correlated reference energy \cite{muolo2018generalized}. 

\begin{figure}[htbp!]
  \centering
  \includegraphics[width = 0.75\linewidth]{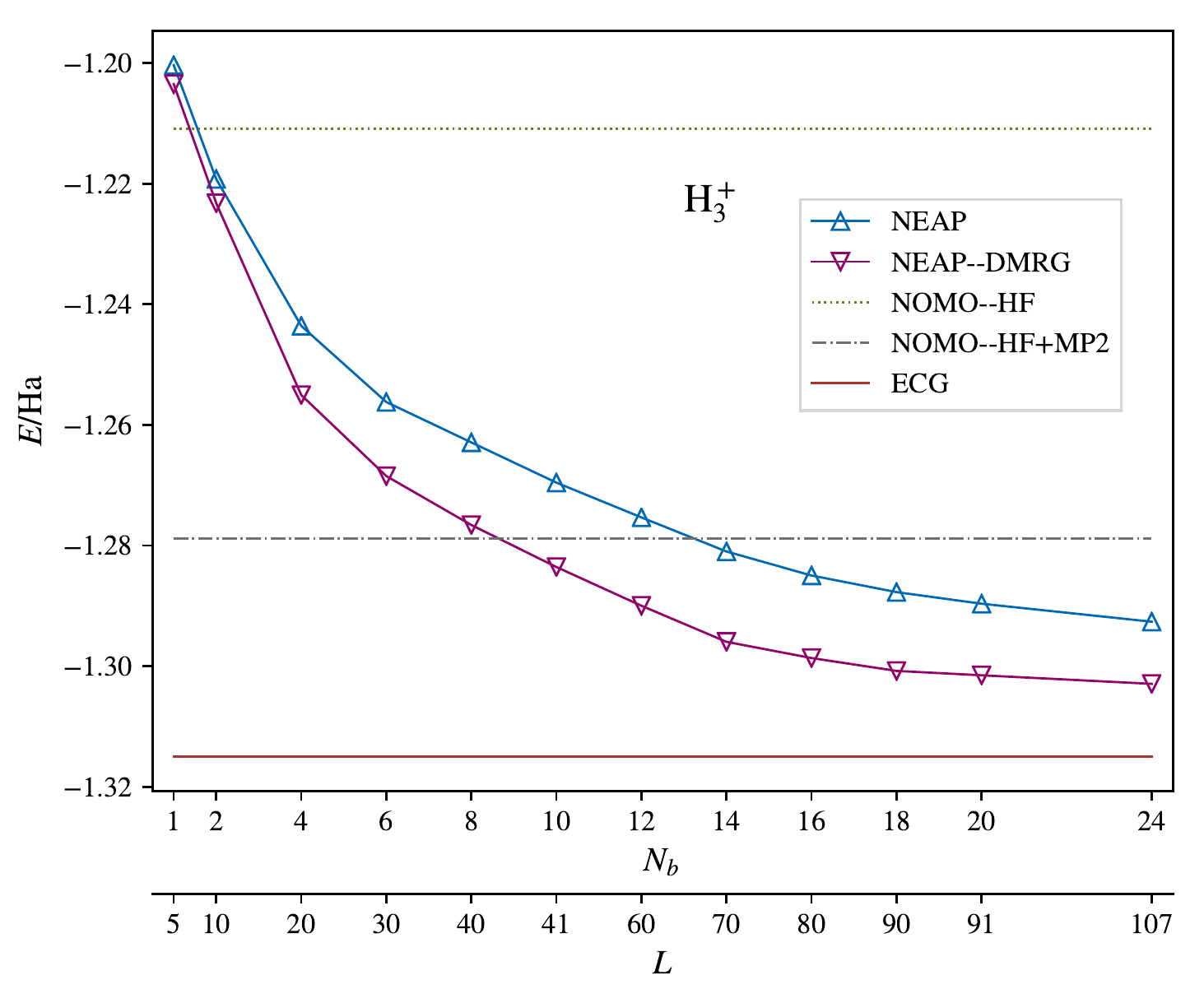}
  \caption{Dependence of the NEAP and NEAP-DMRG ($m=100$) total energy of H$_3^+$ obtained with a $[\text{e}^-(8\text{s})\text{p}^+(3\text{s})]$ basis on the number of functions in the generalized CI expansion ($N_b$) and the dimension of the DMRG lattice ($L$). The horizontal lines refer to energies obtained with nuclear orbital plus molecular orbital Hartree-Fock (NOMO-HF, gray dotted line)~\cite{nakai2007nuclear}, nuclear orbital plus molecular orbital second-order M{\o}ller-Plesset perturbation theory (NOMO--HF+MP2, black double dotted line)~\cite{nakai2007nuclear}, and explicitly correlated Gaussians (ECG, solid red line)~\cite{muolo2018explicitly}. The NEAP--DMRG energy obtained with $N_b$=24 is $-1.30303$ Ha.}
  \label{fig:h3+}
\end{figure}

Next, we apply NEAP-DMRG to the H$_3^+$ molecular ion, for which ECG calculations so far have failed to converge the energy to spectroscopic accuracy \cite{muolo2018explicitly}.
We report in Fig.~\ref{fig:h3+} the NEAP and NEAP-DMRG ground-state energies of H$_3^+$ as a function of $N_b$ and compare our results to NOMO-HF and NOMO-MP2 \cite{nakai2007nuclear} and to explicitly correlated Gaussian \cite{muolo2018explicitly} energies.
Our calculations were carried out with a $[\text{e}^-(8\text{s})\text{p}^+(3\text{s})]$ basis set.
This choice represents a minimal basis for the nuclear part, as at least one primitive Gaussian per nucleus is included, while for electrons a non-minimal primitive basis was chosen, based on the observations on the H$_2$ system.
The lowest NEAP-DMRG energy is by about 0.02 Ha lower than the NOMO-HF+MP2 energy and by about 0.01 Ha higher than the explicitly correlated Gaussian result \cite{muolo2018explicitly}.

Results for the ground state of H$_2$ and H$_3^+$ seem to confirm that single-particle
functions yield a slower energy convergence with respect to the basis set size 
than the explicitly correlated counterparts.
However, while the latter are chosen to be eigenfunctions of both 
the spin and the total spatial angular momentum operators \cite{muolo2018explicitly},
no symmetric wavefunctions are considered in this first version of NEAP-DMRG.
Symmetry adaptation, together with the exploration of larger LCPOs expansions, will be the subject of future work.

\begin{table}[htbp!]
  \centering
  \begin{tabular}{c @{\hspace{4mm}} | @{\hspace{4mm}} c @{\hspace{4mm}} c @{\hspace{4mm}} c @{\hspace{4mm}} c @{\hspace{4mm}} c @{\hspace{4mm}} c}
    \hline\hline
    Threshold  & $10^{-2}$ & $10^{-3}$ & $10^{-4}$ & $10^{-5}$ & $10^{-6}$ & $10^{-10}$ \\
    \hline
    orbitals & $11/6/6$ & $22/13/6$ & $30/14/8$ & $32/15/8$ & $32/20/8$ & $32/24/8$ \\
    \hline 
    $E/$Ha     & $-26.0137$ & $-26.1879$ & $-26.2465$ & $-26.2538$ & $-26.2541$ & $-26.2557$ \\ 
    \hline\hline
  \end{tabular}
  \caption{Ground state total NEAP-DMRG energies of the BH$_3$ molecule obtained with a fixed basis set $\text{e}^-(8\text{s})\text{p}^+(3\text{s})\mathrm{B}^{11}(1\text{s})$ and $N_b=8$, with respect to the L\"owdin threshold and the size of the orbital spaces $\mathcal{B}_i$ (indicated as $\text{dim}(\mathcal{B}_{\text{e}^-})/\text{dim}(\mathcal{B}_{\text{p}^+})/\text{dim}(\mathcal{B}_{\mathrm{B}^{11}})$.}
  \label{tab:thresholdLoewdin}
\end{table}

\begin{figure}[htbp!]
  \centering
  \includegraphics[width = 0.75\linewidth]{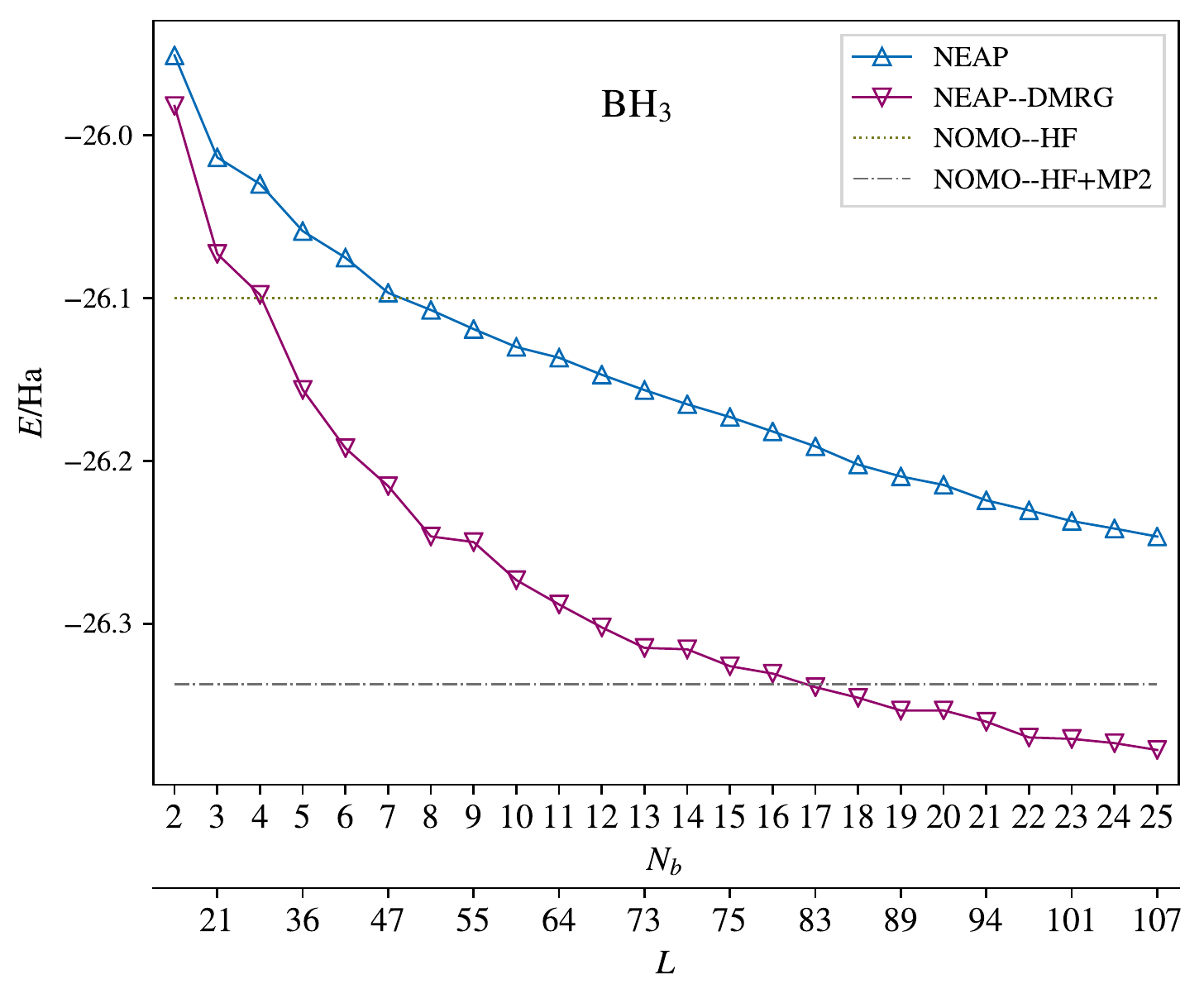}
  \caption{Total NEAP (solid blue line) and NEAP-DMRG (solid purple line) energies of BH$_3$ 
with a $[\text{e}^-(8\text{s})\text{p}^+(1\text{s})\mathrm{B}^{11}(1\text{s})]$ basis and a bond dimension $m$=500 with respect to the number of functions in the generalized CI expansion ($N_b$) and the size of the DMRG lattice ($L$). The horizontal lines refer to nuclear orbital plus molecular orbital Hartree-Fock (NOMO--HF, dotted line) and second-order M{\o}ller-Plessed perturbation theory (NOMO--HF+MP2, double dotted line)~\cite{nakai2007nuclear}. The lowest NEAP--DMRG energy obtained with $N_b$=25 is $-27.36769$~Ha.}
  \label{fig:bh3}
\end{figure}

Finally, as an example that cannot be treated with explicitly correlated wavefunctions because of their factorial scaling, we calculated the ground state of BH$_3$ as a $12$ particle system. 
The calculations were performed with the minimal $[\text{e}^-(8\text{s})\text{p}^+(1\text{s})\mathrm{B}^{11}(1\text{s})]$ basis set. In this minimal basis, the NEAP method reproduces the NOMO-HF reference only for $N_b>8$.
The dependence on the initial guess of the NEAP optimization was ameliorated by considering a pool of 10$^7$ initial trial basis functions from which the best one is permamently included in the basis set.
The number of optimization cycles per basis function spans from 1 million to 100 millions.
The NEAP calculation for BH$_3$ ran on 6 cores of an Intel(R) Xeon(R) Gold 6136 CPU for 3 weeks. The NEAP-DMRG calculation with $N_b=25$ needed 10 days using 16 cores of Intel(R) Xeon(R) CPU E5-2667 v2 CPU.
The reference NOMO-HF calculation employs the cc-pVTZ Dunning basis for the electronic part that includes 90 contracted Cartesian GTOs (16 for each H and 42 for B), whereas the nuclear part includes 63 contracted Cartesian GTOs for each nucleus.
Conversely, the NEAP primitive basis set comprises spherical GTOs: 1 per H nucleus, 1 for the B nucleus and 64 for the electronic part (8 per electron). Therefore, the latter basis is significantly smaller than the HF one and this may be the reason why the NEAP energy obtained with $N_b=1$ is higher than the reference NOMO-HF one.
The DMRG energies are expressed in terms of the maximum lattice length $L$, while the effective lattice length might be smaller in the presence of linearly dependent orbitals.

For a system as large as BH$_3$ the choice of the reduced orthogonal orbital space is decisive in order to save computational time. The arbitrary precision implementation of the orthogonalization step allows us to discard eigenvalues, and corresponding eigenvectors, of the orbital overlap matrix below any user-decided threshold as the error is much below the machine accuracy of double precision arithmetic.
We highlight that, as the Fock space dimension is related to the system size and to the orbital basis size through the binomial relation of Eq.~(\ref{eq:DimensionManyBodySpace}), the number of multicomponent configurations spanned by NEAP-DMRG increases more rapidly with $N_b$ for BH$_3$ than for H$_2$ or H$_3^+$. Hence, the energy difference between NEAP of NEAP-DMRG is increasing.

In Table~\ref{tab:thresholdLoewdin} we study the effect of discarding an increasing number of orbitals by progressively decreasing the L\"owdin threshold. 
We find that a L\"owdin threshold of $10^{-4}$ delivers a good compromise between the energy and the number of orbitals, and hence the DMRG computational cost.
Assuming that the difference displayed for $N_b=8$ does not increase for higher $N_b$, we consider this threshold hereafter.
We studied energy convergence of the ground state of BH$_3$ for a number of stochastically optimized multicomponent orbitals by discarding overlap eigenvalues lower than $10^{-4}$ and with LCPO sizes being $\text{e}^-(8\text{s})\text{p}^+(3\text{s})\mathrm{B}^{11}(1\text{s})$. 
In Fig.~\ref{fig:bh3}, we compare NEAP and NEAP-DMRG results with NOMO-HF and NOMO-MP2 results taken from Ref. \cite{nakai2007nuclear}.
Our best variational upper bound obtained with NEAP-DMRG is $E_{\text{BH}_3}=-26.37769$~Ha. Compared with earlier results, this is $0.27782$~Ha lower than NOMO-HF and $0.04037$~Ha lower than NOMO-MP2 for the same translational-free Hamiltonian \cite{nakai2007nuclear}.
We collect in Table~II the best variational upper bound to the energy for H$_2$, H$_3^+$, and BH$_3$ obtained with NEAP and NEAP-DMRG.

\begin{table}[htbp!]
  \centering
  \begin{tabular}
    {l @{\hspace{5mm}} l @{\hspace{6mm}} c @{\hspace{6mm}} c @{\hspace{6mm}} c @{\hspace{6mm}} c}
    \hline\hline
    System      & Basis set                      & $N_b$ & $m$ & $E_{\mathrm{NEAP}}$ & $E_{\mathrm{DMRG}}$ \\
    \hline
    H$_2$       & $e^-(6s)p^+(2s)$               & $32$  & $60$  & $-1.14393$     & $-1.15226$  \\ 
    H$_3^+$     & $e^-(8s)p^+(3s)$               & $24$  & $100$ & $-1.29208$     & $-1.30303$  \\
    BH$_3$      & $e^-(8s)p^+(3s)\mathrm{B}(1s)$ & $25$  & $500$ & $-26.246529$    & $-26.37769$ \\
    \hline\hline
  \end{tabular}
  \caption{NEAP and NEAP-DMRG ground-state energy (in Hartree atomic units) of H$_2$, H$_3^+$, and BH$_3$ obtained with the largest values of the size of the generalized CI expansion ($N_b$) and bond dimension ($m$).}
  \label{tab:my_label}
\end{table}

\section{Conclusions}
\label{sec:conclusions}

In this work, we introduce the \textit{ab initio} NEAP-DMRG approach to determine fully coupled
electronic and nuclear wave functions without relying on the BO approximation.
We introduce a multiconfigurational ansatz for the total molecular wave function and expand both the nuclear and electronic orbitals as a linear combination primitive spherical Gaussians.
The flexibility of the ansatz is enhanced by considering non-orthogonal orbitals and by optimizing all Gaussian parameters, \textit{i.e.}, widths and shifts, variationally.
In this respect, NEAP-DMRG differs from most multicomponent orbital-based methods available in literature that rely on molecular orbitals constructed from a mean-field calculation, such as Hartree--Fock.
However, due to the strong electron-nuclear correlation, HF orbitals are known to be an inaccurate reference for a multicomponent post-HF treatment \cite{HammesSchiffer2004_Tunneling-Correlation}.

\begin{figure}[htbp!]
  \centering
  \includegraphics[width=0.9\linewidth]{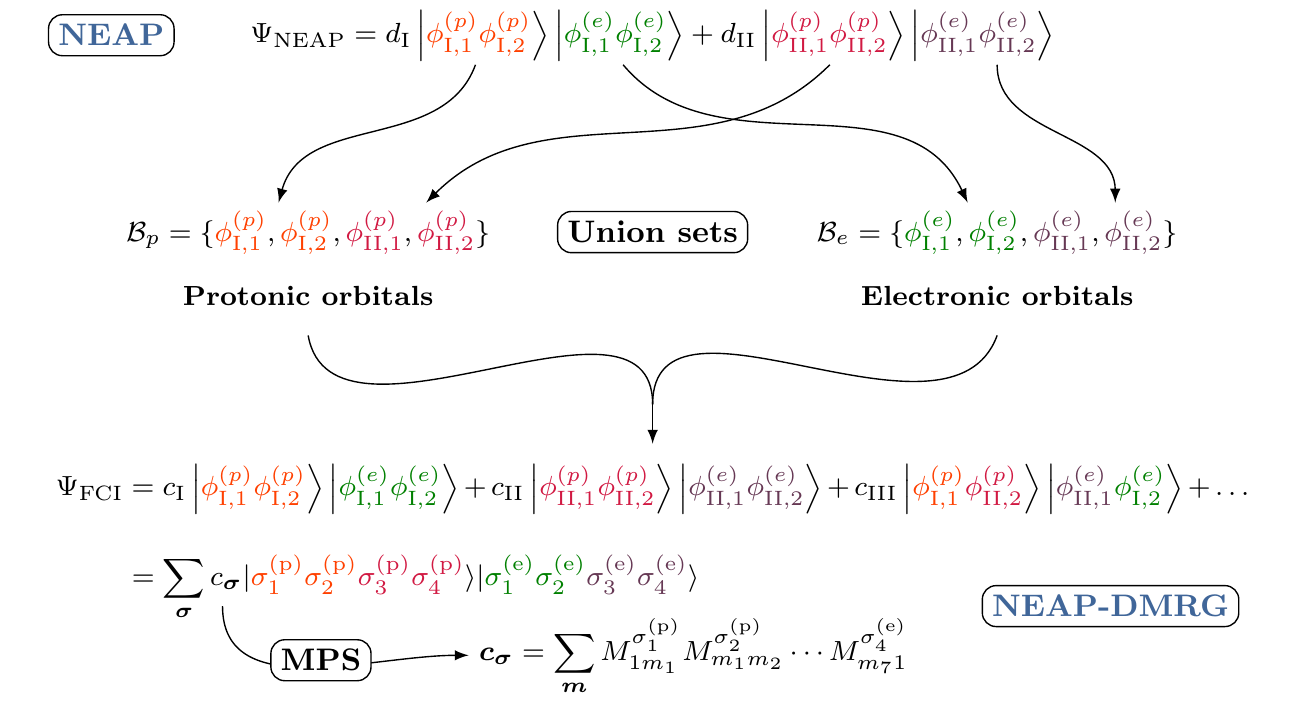}
  \caption{Schematic definition of the NEAP-DMRG wave function for the H$_2$=$\{$e$^-,$e$^-,$p$^+,$p$^+\}$ molecular system in terms of the non-orthogonal stocastically optimized NEAP orbitals. A generalized CI expansion with $N_b$=2 (top of the figure) is considered to define the orbital union sets for each particle type. A FCI expansion is defined in the basis of these union sets and then expressed as an MPS (bottom of the figure).}
  \label{fig:Workflow}
\end{figure}

The stochastic optimization is based on the competitive selection scheme and yields the optimal set of orbitals for the multiconfigurational trial wave function.
Based on this optimal orbital set, we construct a generalized CI wave function and optimize it by applying the DMRG algorithm to avoid the exponential scaling of the computational cost.
We work out the extension of DMRG in the MPO/MPS-based formulation to full molecular Hamiltonians comprising different quantum species \cite{Dolfi2014_ALPSProject,keller2015efficient,Baiardi2017_VDMRG}.
We illustrate in Figure~\ref{fig:Workflow} how the generalized CI wave function is constructed from the NEAP non-ortogonal orbitals for a small system,  H$_2$ with $N_b=2$.

NEAP-DMRG yields energies of H$_2$ and H$_3^+$ beyond those of multicomponent approaches based on the HF-CISD(T) or HF-MP2 methods.
Numerical results on BH$_3$ described as an explicit 12-particle system show that DMRG significantly lowers the total energies of multireference trial wave functions.
This demonstrates that DMRG can significantly extend the range of applicability of orbital-based multicomponent schemes.

\clearpage

\section*{Acknowledgments}

This work was supported by ETH Zurich through the ETH Fellowship No. FEL-49 18-1.

%

\end{document}